\documentclass[a4paper,fleqn,usenatbib,useAMS]{mnras}

\usepackage{graphicx}	
\usepackage{amsmath}	
\usepackage{amssymb}	
\usepackage{multicol}        
\usepackage{bm}		
\usepackage{pdflscape}	
\usepackage{dcolumn}

\newcommand{\msol}{M_{\rm \odot}} 
\newcommand{\mjup}{M_{\rm Jup}} 

\newcommand{\white}[1]{\textcolor{white}{#1}}

\newcolumntype{d}{D{.}{.}{2.5}}
\newcolumntype{s}{D{.}{.}{1.3}}

\usepackage[english]{babel}
\usepackage{blindtext}

\usepackage[T1]{fontenc}
\usepackage{ae,aecompl}

\usepackage{newtxtext,newtxmath}

\title[Chemistry of fragments formed via GI]{The chemistry of protoplanetary fragments formed via gravitational instabilities}

\author[J.~D.~Ilee et al.]{%
J.~D.~Ilee$^{1}$\thanks{Contact e-mail: \href{mailto:jdilee@ast.cam.ac.uk}{jdilee@ast.cam.ac.uk}},
D.~H.~Forgan$^{2,3}$,
M.~G.~Evans$^{4}$,
C.~Hall$^{5,6}$,
R.~Booth$^{1}$,
C.~J.~Clarke$^{1}$, 
\newauthor
W.~K.~M.~Rice$^{5,7}$, 
A.~C.~Boley$^{8}$,
P.~Caselli$^{9}$,
T.~W.~Hartquist$^{4}$
and J.~M.~C.~Rawlings$^{10}$
 \vspace{0.4em}
\\
$^{1}$Institute of Astronomy, Madingley Road, Cambridge CB3 0HA, UK\\
$^{2}$SUPA, School of Physics \& Astronomy, University of St Andrews, North Haugh, St Andrews, Scotland, KY16 9SS, UK\\
$^{3}$Centre for Exoplanet Science, University of St Andrews, St Andrews, UK \\
$^{4}$School of Physics and Astronomy, University of Leeds, Leeds LS2 9JT, UK\\
$^{5}$SUPA, Institute for Astronomy, University of Edinburgh, Blackford Hill, Edinburgh EH9 3HJ, UK\\
$^{6}$Department of Physics \& Astronomy, University of Leicester, University Road, Leicester, LE1 7RH, UK\\
$^{7}$Centre for Exoplanet Science, University of Edinburgh, Edinburgh, UK\\
$^{8}$Department of Physics and Astronomy, The University of British Columbia, 6224 Agricultural Road, Vancouver V6T 1Z4, Canada\\
$^{9}$Max-Planck Institute for Extraterrestrial Physics, Giessenbachstrasse 1, D-85748 Garching, Germany \\
$^{10}$Department of Physics and Astronomy, University College London, Gower Street, London WC1E 6BT, UK \\
}

\date{Accepted 2017 July 27. Received 2017 July 27; in original form 2017 June 23.}

\pubyear{2017}

\begin{document}
\label{firstpage}
\pagerange{\pageref{firstpage}--\pageref{lastpage}}
\maketitle

\begin{abstract}
In this paper, we model the chemical evolution of a 0.25~$\msol$ protoplanetary disc surrounding a 1~$\msol$ star that undergoes fragmentation due to self-gravity.  We use Smoothed Particle Hydrodynamics including a radiative transfer scheme, along with time-dependent chemical evolution code to follow the composition of the disc and resulting fragments over approximately 4000\,yrs.  Initially, four quasi-stable fragments are formed, of which two are eventually disrupted by tidal torques in the disc.  From the results of our chemical modelling, we identify species that are abundant in the fragments (e.g. H$_{2}$O, H$_{2}$S, HNO, N$_{2}$, NH$_{3}$, OCS, SO), species that are abundant in the spiral shocks within the disc  (e.g. CO, CH$_{4}$, CN, CS, H$_{2}$CO), and species which are abundant in the circumfragmentary material (e.g. HCO$^{+}$).  Our models suggest that in some fragments it is plausible for grains to sediment to the core before releasing their volatiles into the planetary envelope, leading to changes in, e.g., the C/O ratio of the gas and ice components.  We would therefore predict that the atmospheric composition of planets generated by gravitational instability should not necessarily follow the bulk chemical composition of the local disc material. 
\end{abstract}




\section{Introduction}

The star formation process results in the production of circumstellar discs around young protostars.  These discs provide further feedstock for the star to continue accreting material, but they are also the sites of planet formation.  Planet formation models can be split into two camps: the bottom-up process of core accretion (CA, \citealt{pollack_1996}), which relies on the coagulation of dust grains into larger bodies, or via the top down process of disc fragmentation via the gravitational instability (GI, \citealt{durisen_2007, helled_2014}).  

\smallskip

\begin{figure*}
\includegraphics[width=0.49\textwidth]{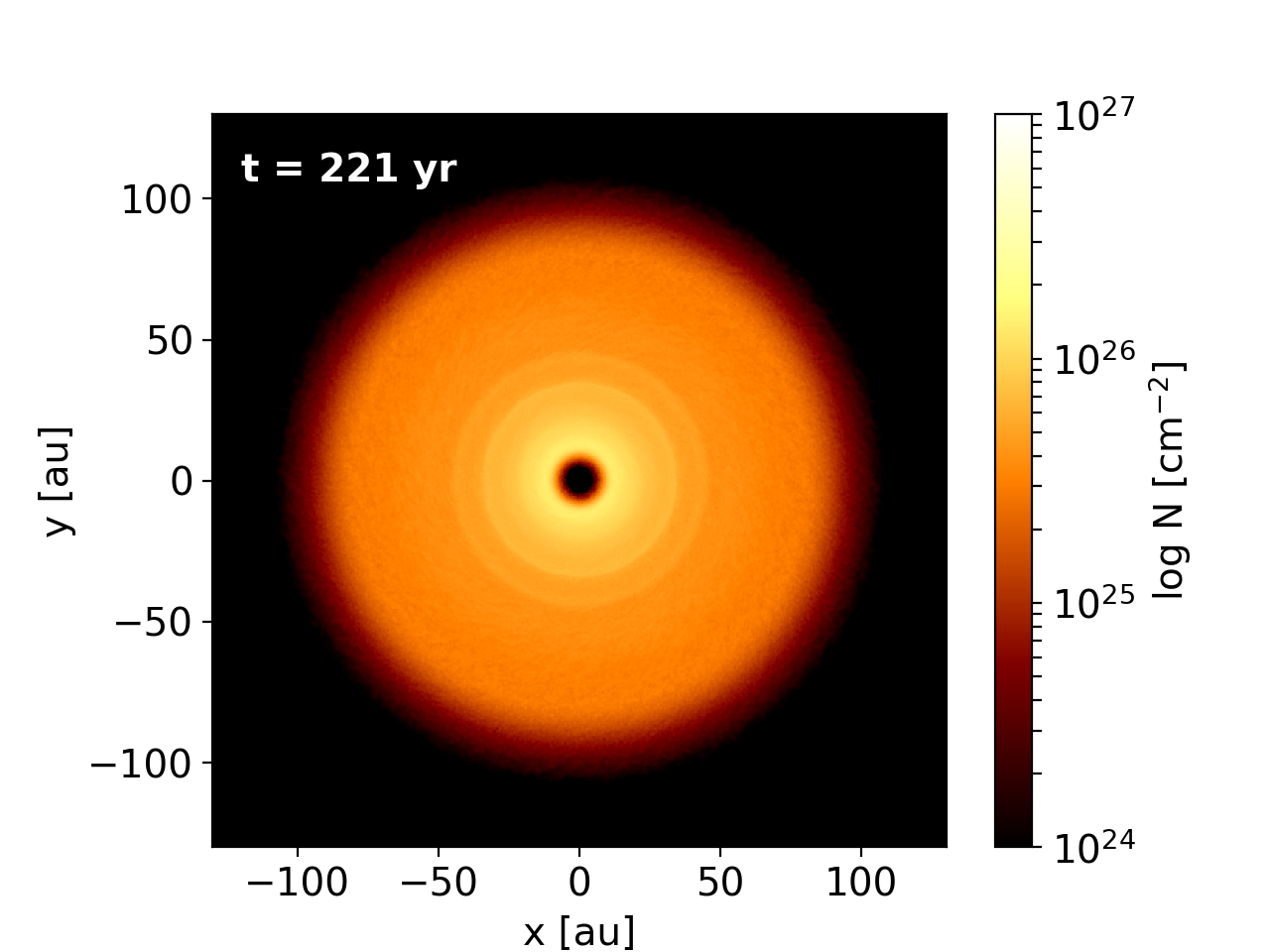}
\includegraphics[width=0.49\textwidth]{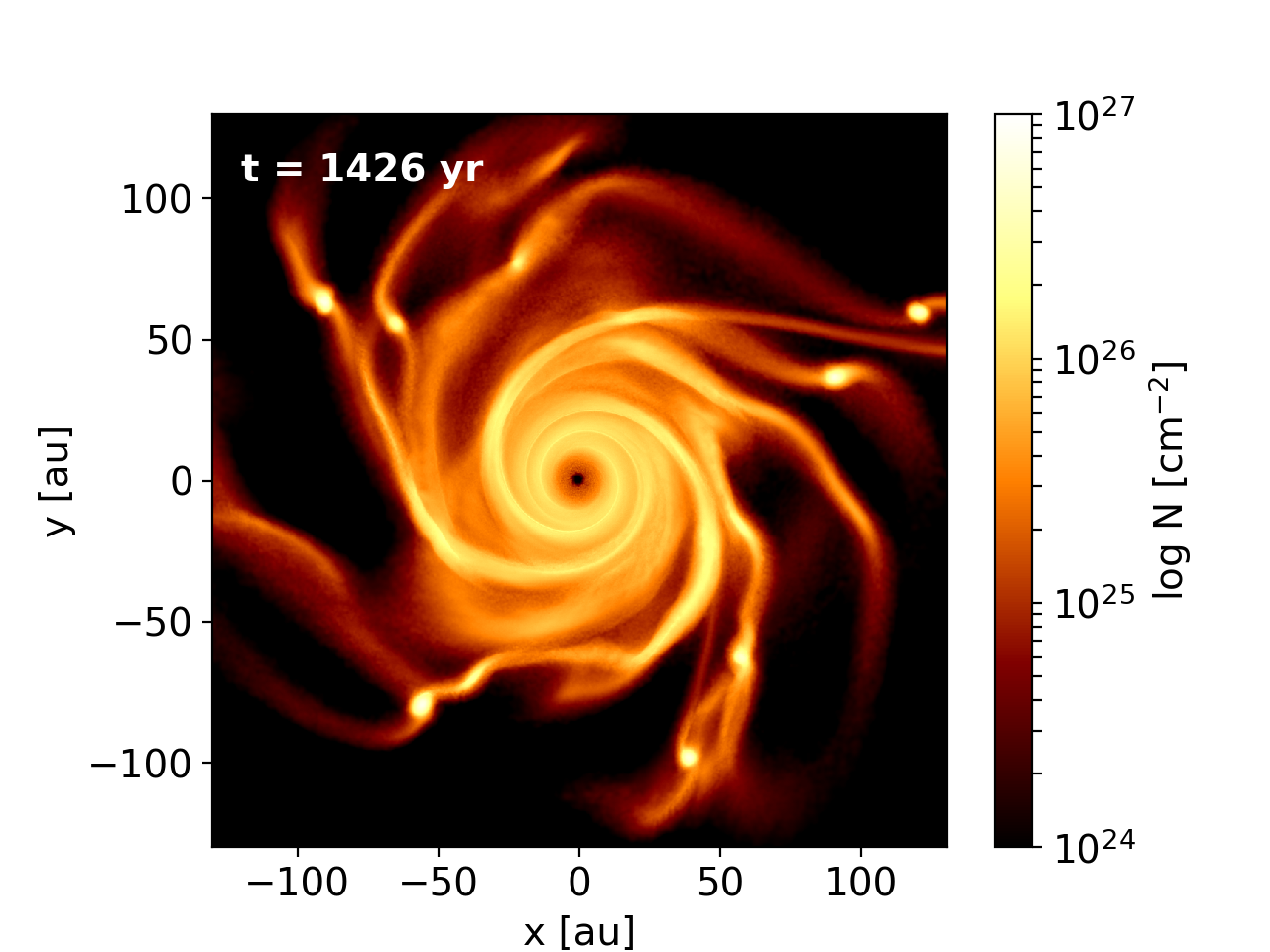}\\
\includegraphics[width=0.49\textwidth]{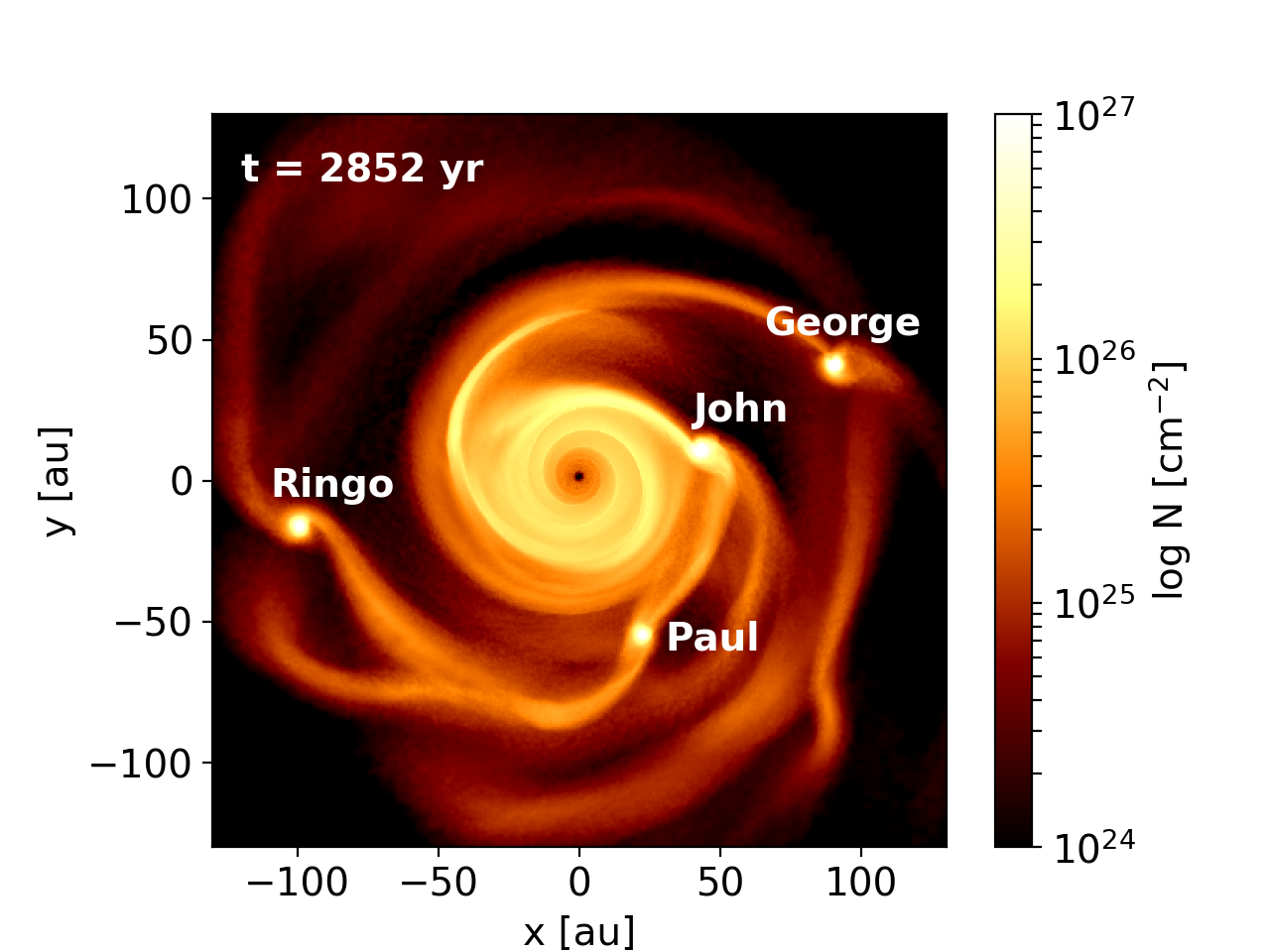} 
\includegraphics[width=0.49\textwidth]{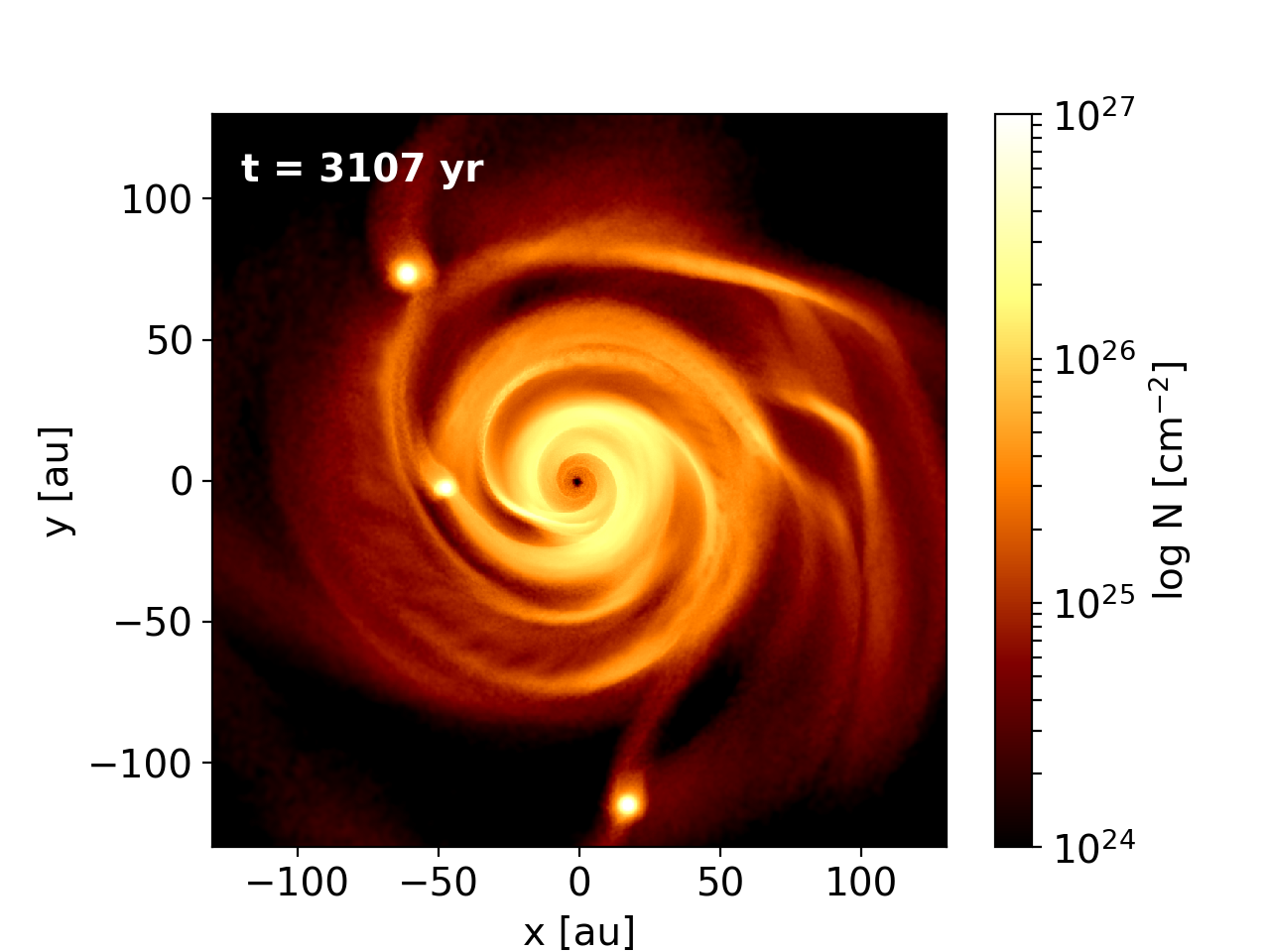}\\
\includegraphics[width=0.49\textwidth]{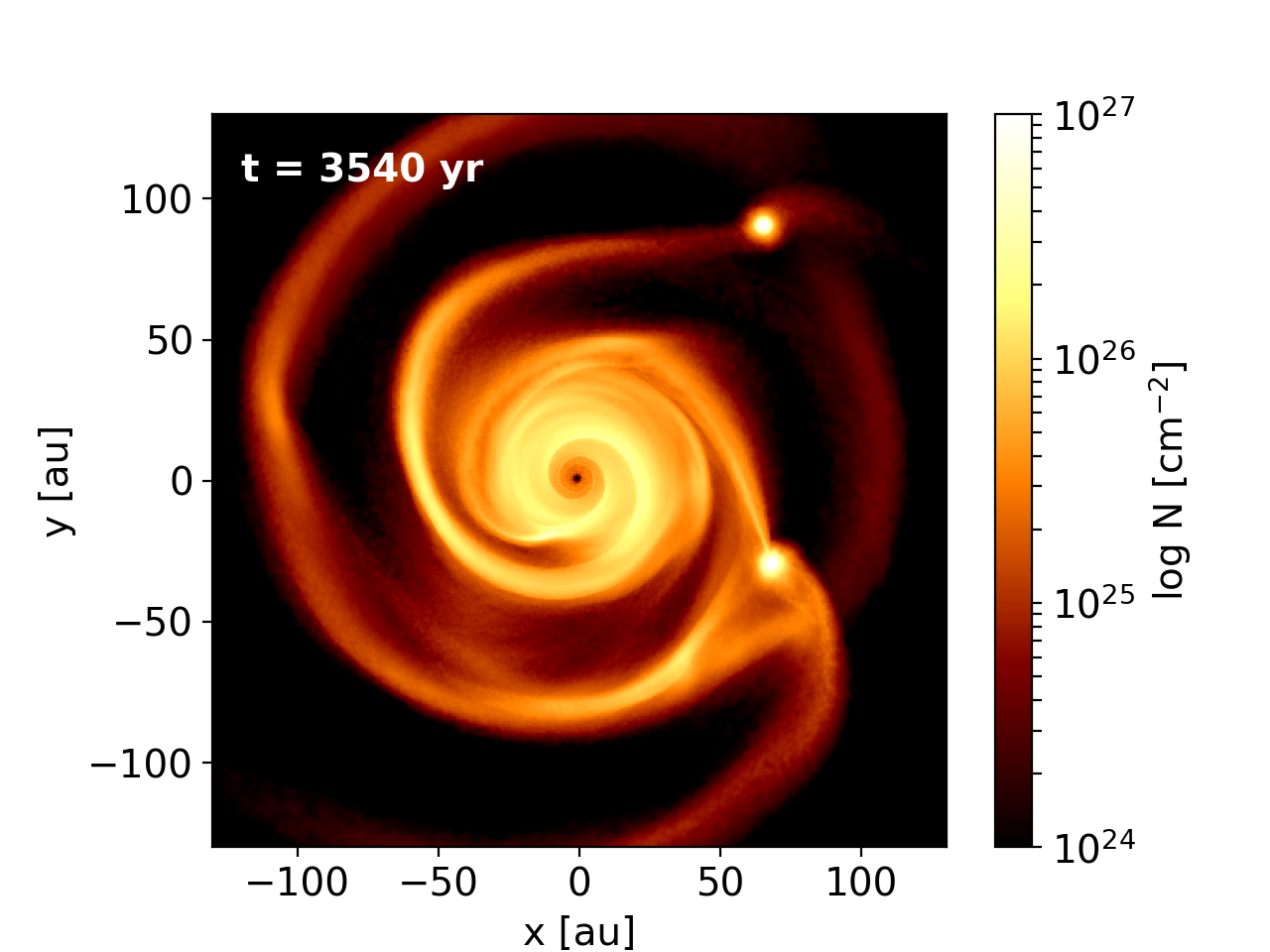}
\includegraphics[width=0.49\textwidth]{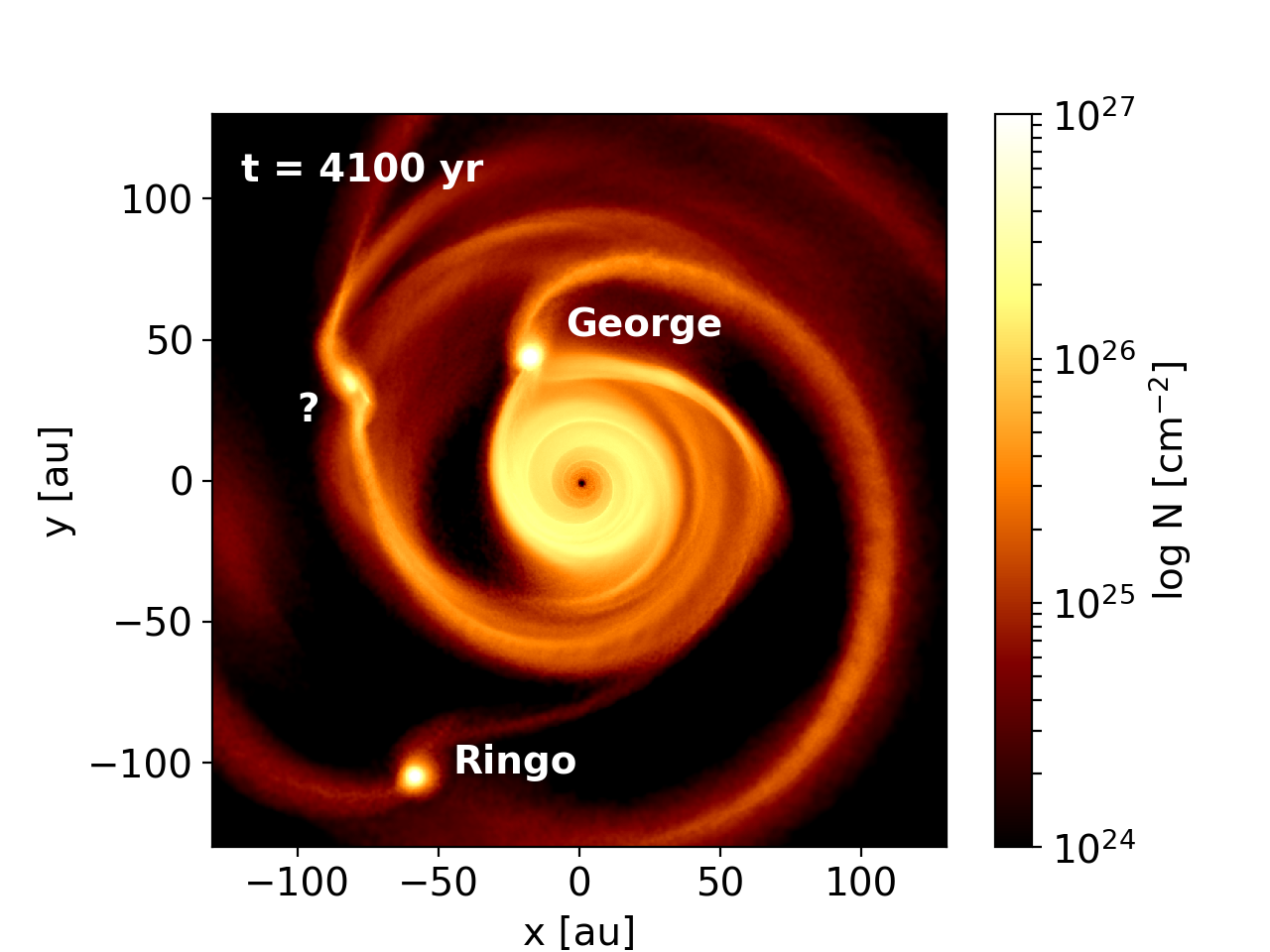}\\
\caption{The evolution of the hydrodynamic simulation.  The snapshots cover approximately 4000~yrs of disc evolution, through fragmentation into a number of objects, of which four are able to resist shearing and spiral arm shocks, which we label John, Paul, George and Ringo in the middle left panel.  John and Paul are tidally disrupted (middle right and bottom left panels), and a fifth fragment appears to be formed via triggering in the final snapshot (bottom left).}
\label{fig:hydro_evolution}
\end{figure*}

\begin{figure*}
\includegraphics[width=0.49\textwidth]{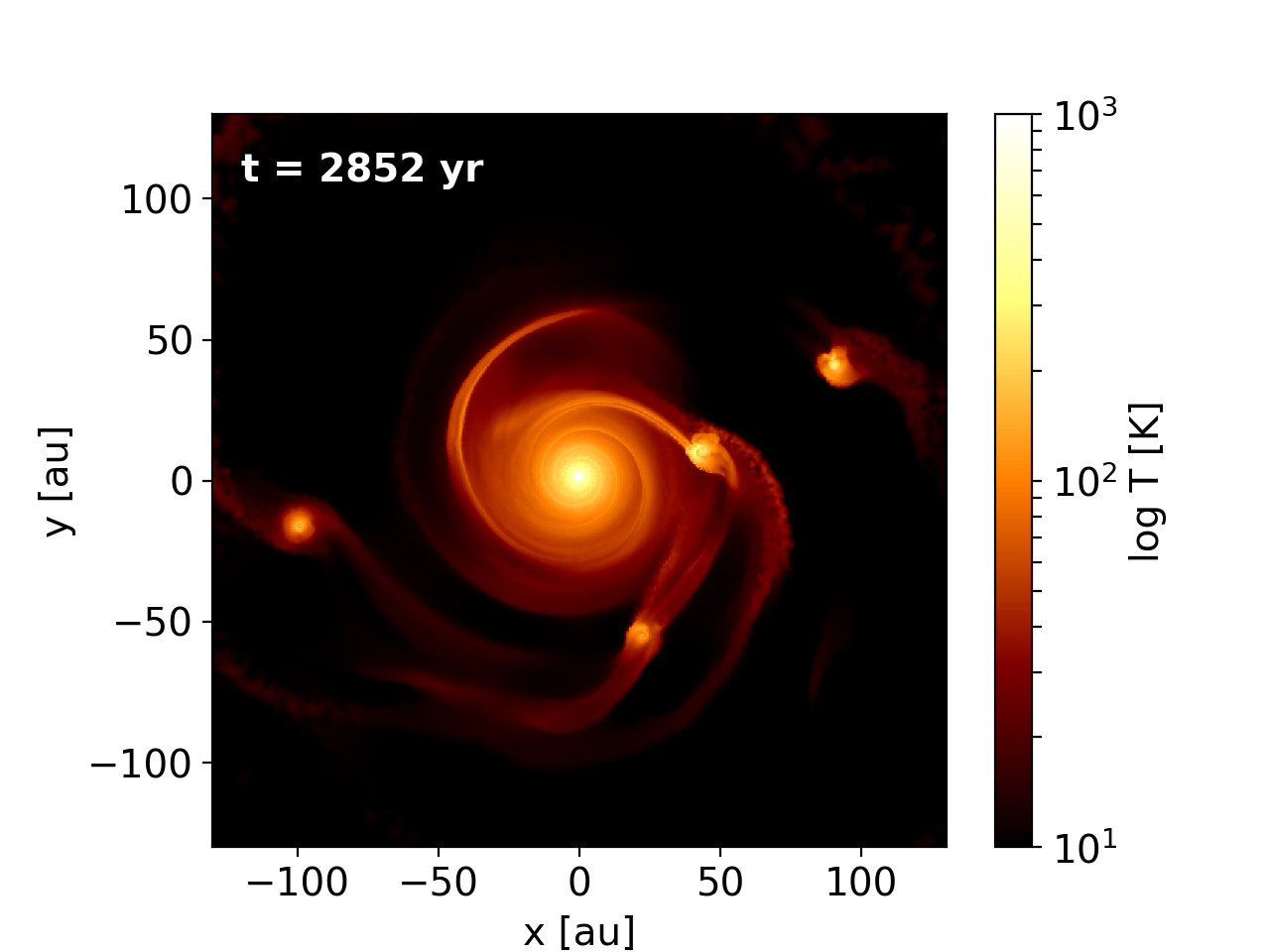}
\includegraphics[width=0.49\textwidth]{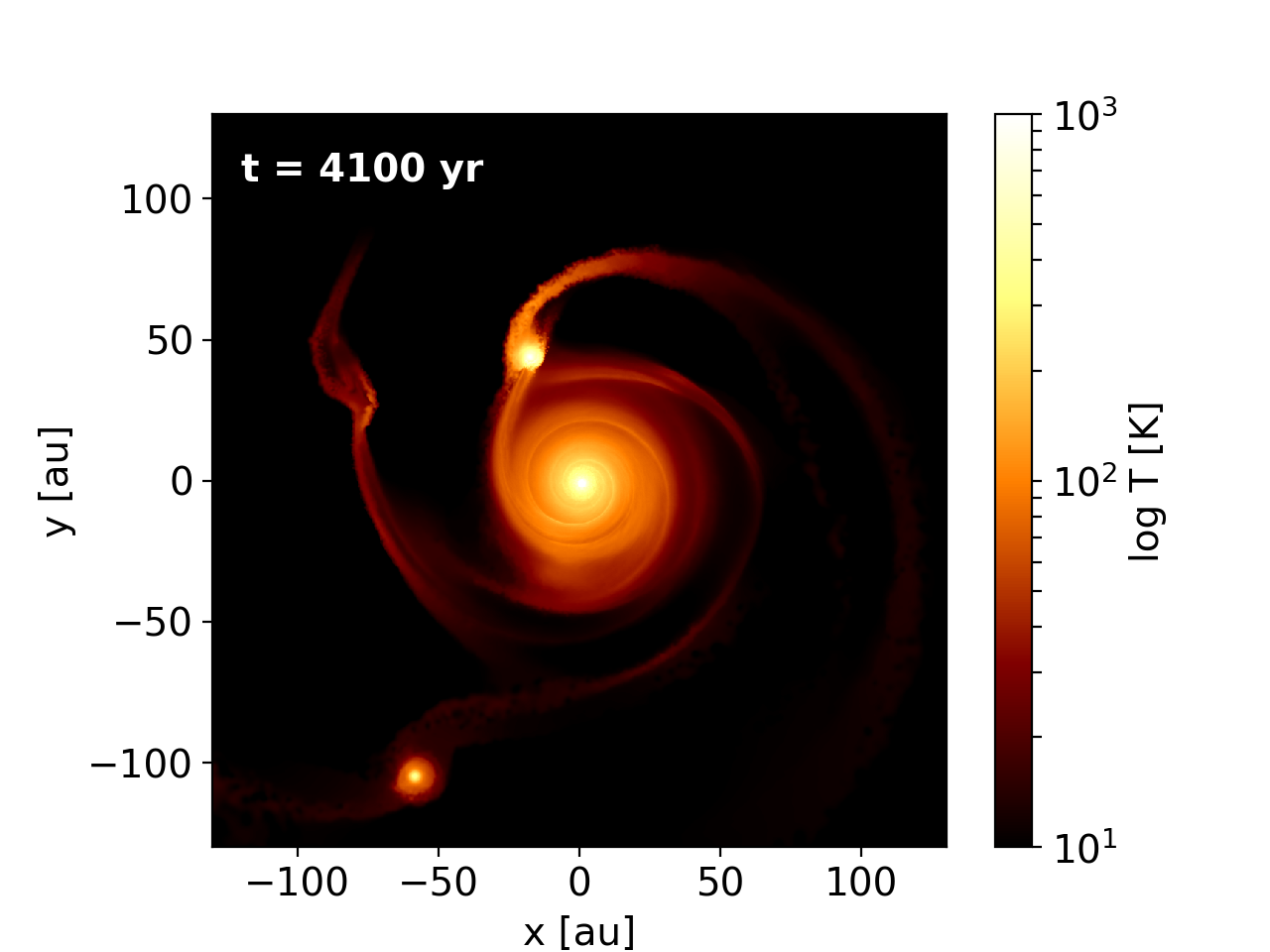} 
\caption{Mid-plane temperatures of the disc simulation at $t=2852$ (left) and 3913 yrs (right), corresponding to the third and sixth panels Figure \ref{fig:hydro_evolution}.  Globally, the disc exhibits an expected radially-decreasing temperature gradient, but significant increases in temperature are seen around the fragments and spiral arms.}
\label{fig:temperature}
\end{figure*}

At the birth of the star-disc system, it is expected that the disc mass is approximately equal to the mass of the star, and observations constrain the disc-to-star mass ratios to be no less than 0.5 \citep[e.g.][]{eisner_2006,    joos_2012}.  At such high disc-to-star mass ratios, the evolution of the system is likely to be dominated by disc self-gravity. These discs will exist in a state of marginal instability if the  local Toomre Parameter \citep{toomre_1964},
\begin{equation} 
Q = \frac{c_{s}\kappa}{\pi G \Sigma} \sim 1, 
\end{equation}
where $\kappa$ is the epicyclic frequency (representing the rotational support, and  equal to  the angular velocity  $\Omega$ in  a Keplerian disc),  $c_{s}$  is the  sound  speed  of  the gas  (representing  the pressure support) and  $\Sigma$ is the surface density  of the gas.  The combined action of thermodynamics, self-gravity and rotation can result in self-regulated, quasi-steady discs where $Q\sim 1.5-1.7$, which is approximately the critical value for non-axisymmetric instabilities \citep{durisen_2007}.  Such discs possess large scale spiral density waves, and these (combined with the gravito-turbulence produced by their non-linear coupling,   \citealt{gammie_2001}) are efficient outward transporters of angular momentum, and can provide a mechanism for driving mass accretion during the earliest stages of star formation.  As as a result, the disc material accretes rapidly onto the central star, in a process that can last for approximately $10^5$ years \citep{lodato_2004, rice_2010}.

\smallskip

Self-gravitating discs may also undergo fragmentation if $Q\sim 1$ and the surface density perturbations induced by the gravitational instability are massive enough, and their cooling is efficient enough such that they become gravitationally self-bound and collapse \citep{gammie_2001, rice_2005, meru_2011, paardekooper_2012, young_2016}.  The process of disc fragmentation has largely been the province of theory, but several works have shown that the Atacama Large Millimetre Array (ALMA) has both the high angular resolution and sensitivity to directly detect gravitationally unstable discs in the nearest star forming regions \citep[e.g.][]{cossins_2010, douglas_2013, dipierro_2014, dipierro_2015, hall_2016, evans_2017}.  Indeed, recent observational campaigns are beginning to reveal spiral arm structures at sub-millimetre wavelengths \citep[][]{Perez_2016}, a hallmark of GI or recent fragmentation \citep[][]{Meru_2017}.  Most strikingly, evidence of a fragmenting disc was gathered by \citet{Tobin_2016}, showing a star-disc system with two stellar mass objects in the disc connected by spiral structure.  This system demonstrates that GI and disc fragmentation can form stellar mass objects.  But can this process form planetary mass objects?

\smallskip

A standing assumption is that, in contrast to protoplanets formed via CA, a fragment formed via GI would inherit an initial composition that is likely to reflect the bulk stellar or bulk disc composition.  However, \citet{boley_2010a} and \citet{boley_2011} discuss several situations for which this assumption may not hold.  Firstly, if a population of grains is present at the birth of the disc, then grains with stopping distances comparable to the widths of the spiral arms will be efficiently collected in the arms prior to fragmentation.  Subsequent fragmentation of this arm would result in fragments with a super-stellar metallicity.  Secondly, if for some reason a population of planetesimals have formed prior to fragmentation, then the gas will be inefficient at damping their motion.  This effect will persist, even within a fragment, until temperatures reach the H$_{2}$ dissociation limit.  Thus, the planetesimal  concentration will be lower than the fragment concentration for some time, leading to sub-stellar metallicities.  Thirdly, within a fragment itself, the initial sample of dust grains will begin to collide and grow in a very similar fashion to grain growth as predicted by CA, as well as accreting further solid material as it traverses through the circumstellar disc \citep{helled_2006}.  If grains can grow to sufficiently large sizes, they will sediment towards the pressure maximum at the centre of the fragment, and in the right circumstances can form a solid core. Finally, planet formation depletes metals from the gas-phase.  If this metal-depleted gas continues to accrete onto the central star, then the stellar atmospheric composition may diverge from the bulk composition of the initial disc, leading to sub-stellar metallicities.  Such depletions are seen toward stars exhibiting the $\lambda$-Bootis phenomenon \citep[e.g.][]{heiter_2002}. 

\smallskip

The combined action of these processes results in a much wider variety of objects formed via GI than was previously considered, and has resulted in a revamped model of fragmentation usually referred to as `tidal downsizing' \citep{boley_2010,nayakshin_2010a}.  It has been suggested that giant planets form with a variety of core masses under this paradigm \citep{nayakshin_2015}, and that the typical core mass is larger than previously predicted (e.g. \citealt{saumon_2004}).  However, population synthesis models indicate that the formation of terrestrial planets via solid core formation and disruption of the outer gaseous envelope is an unlikely outcome \citep{forgan_2013}.

\smallskip

The majority of GI objects appear to be giant planets and brown dwarfs, both bound to stars and free-floating \citep{forgan_2015}.  Observational constraints \citep{Vigan2017} suggest that such objects are rare compared to GI predictions, indicating that the proportion of exoplanets that are GI objects is potentially small.  However, there are examples of planetary systems that include giant objects on wide orbits \citep[e.g. HR 8799,][]{marois_2008}, or systems which are undergoing or might have undergone fragmentation \citep[e.g. Elias 2-27,][]{Meru_2017}, suggesting it is possible for GI to form some planetary systems.  In order to provide further observational predictions that can be used to examine the properties of these planets, and the discs that formed them, it is important to consider the chemical composition of both.  

\smallskip

Chemical evolution in protoplanetary discs has been extensively studied \citep[see the reviews by, e.g.,][]{bergin_2007,henning_2013,dutrey_2014}.  Many of the models to date have focused on immediately observable characteristics, such as species responsible for emission lines formed at high relative disc heights.  However, there has recently been more interest in evaluating the chemical properties of the bulk of the planet building material in the midplane \citep[e.g.][]{eistrup_2016}.  Unfortunately, due to the computationally demanding nature of \emph{i)} evolving large chemical networks of many thousands of species and/or \emph{ii)} calculating the mutual influences of chemistry on the thermal evolution of the disc, these models are at most two-dimensional in nature \citep[see the review of][]{haworth_2016}.  Such a setup limits their ability to investigate the chemical evolution of structures that are non-axisymmetric in nature.  There are a handful of studies concerned with investigating chemical evolution in three dimensional disc models -- e.g. young, massive discs with non-axisymmetric structure \citep{ilee_2011, hincelin_2013, evans_2015}.  However, these studies either included discs that did not fragment, or did not follow the physical evolution for a sufficiently long time to characterise any fragments that may be formed within the discs.

\smallskip

In this paper, we use a Smoothed Particle Hydrodynamics (SPH) model with a hybrid radiative transfer scheme along with non-equilibrium gas-grain chemical evolution code to model the evolution of a fragmenting, self-gravitating disc.  We characterise both the effect of the dynamical evolution on the chemical evolution of the disc, and the chemical properties of the resulting fragments themselves.  Section \ref{sec:modelling} outlines our physical and chemical modelling procedures, Section \ref{sec:results} presents our results, Section \ref{sec:discussion} discusses the implications of our work and finally our conclusions are presented in Section \ref{sec:conclusions}.

\section{Modelling Approach}
\label{sec:modelling}

\subsection{Hydrodynamics}

We use the smoothed particle hydrodynamics (SPH) code \texttt{sphNG} \citep{bate_1995} which was modified to include the hybrid radiative transfer formalism of \citet{forgan_2009}. The gas is able to cool radiatively according to its local optical depth, which is estimated from the local gravitational potential \citep{stamatellos_2007a}, and is also able to exchange energy with neighbouring fluid elements via flux-limited diffusion \citep[see e.g.][]{bodenheimer_1990,whitehouse_2004,mayer_2007}.  This approximation allows high resolution simulations to be run with a minimal computational penalty.  Tests show this algorithm increases runtime by only 6 per cent over non-radiative simulations, while still passing key tests of grey radiative transfer algorithms.

\smallskip

We simulate a 0.25~$\msol$ disc, orbiting a 1~$\msol$ star.  The gas in the disc is represented by $4\times10^{6}$ SPH particles distributed between 10 and 100~au, with a surface density profile that scales as $\Sigma (r) \propto r^{-1}$ and a sound speed profile that scales as $c_{\rm s} \propto r^{-0.5}$.  The initial conditions are such that $Q$ decreases with radius so that the value at the outer edge of the disc is approximately 1.  Once the simulation begins, $Q$ evolves to an approximately constant value of 1 between 50 and 90\,au, before fragmentation ensues.  These parameters are consistent with observations of disc-like structures in Keplerian rotation around Class 0 objects \citep[e.g.][]{tobin_2015},  and the disc-star mass ratio is consistent with that expected after the initial rapid accretion of disc material before the envelope dissipates \citep{forgan_2011}.  

\smallskip

The star is represented by a sink particle which can accrete gas particles if they approach sufficiently closely (within 1~au) and are gravitationally bound \citep{bate_1995}.  The creation of additional sink particles is not allowed in the model.  This enables us to follow the formation and evolution of any fragments, but in turn limits the maximum time for which the simulation can be run ($t=4100$\,yr) since beyond this point, densities within the fragments become too large to follow without an unrealistically small hydrodynamic time step.  

\begin{figure*}
\includegraphics[width=0.49\textwidth]{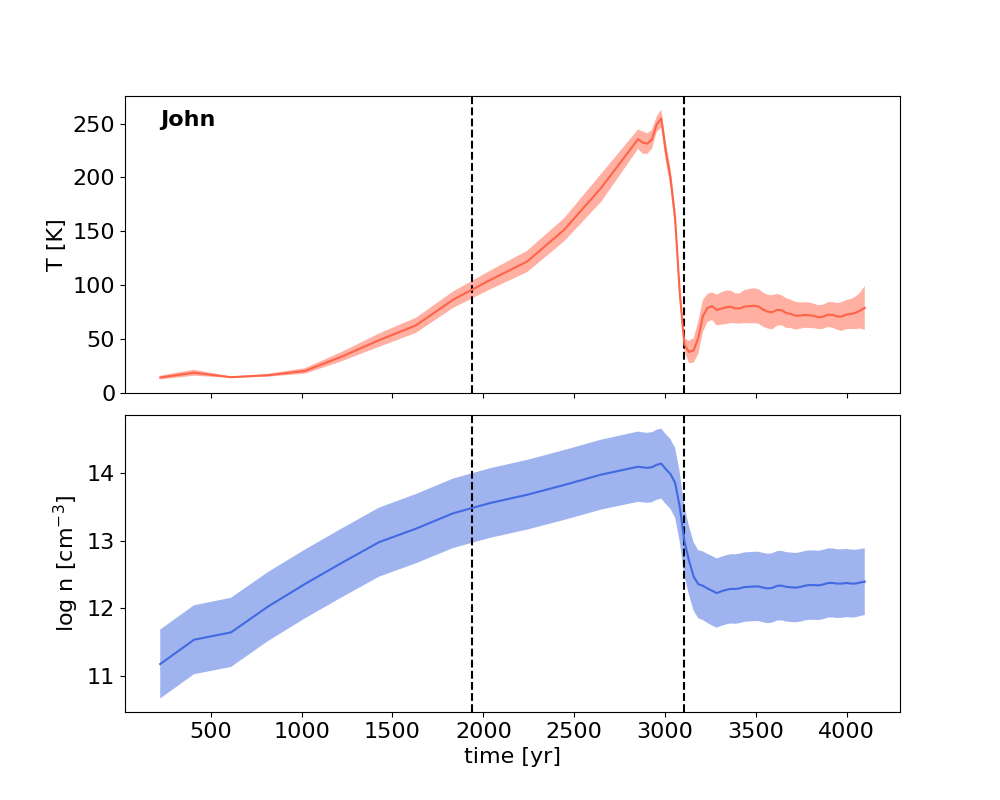}
\includegraphics[width=0.49\textwidth]{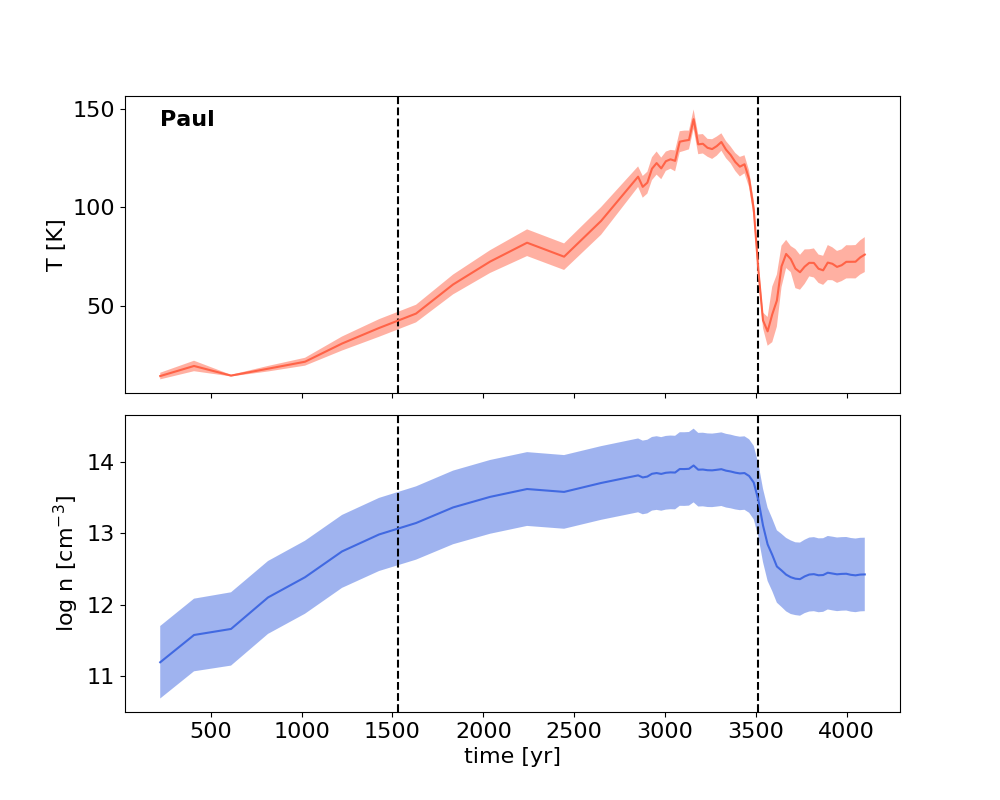}\\

\includegraphics[width=0.49\textwidth]{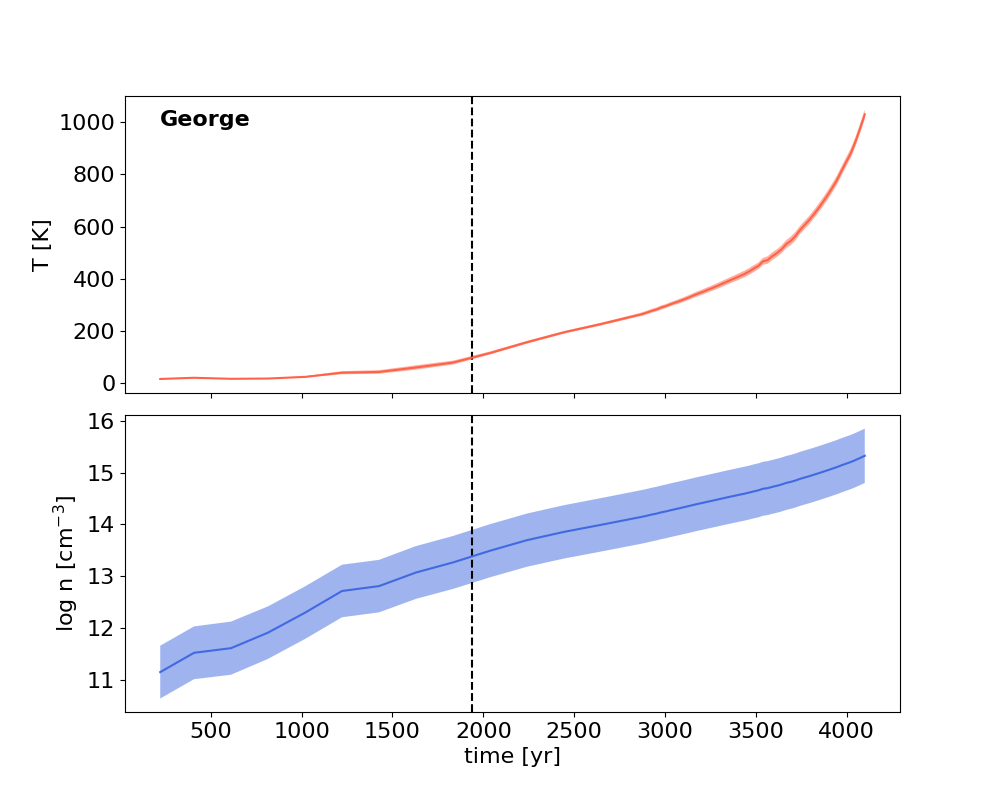}
\includegraphics[width=0.49\textwidth]{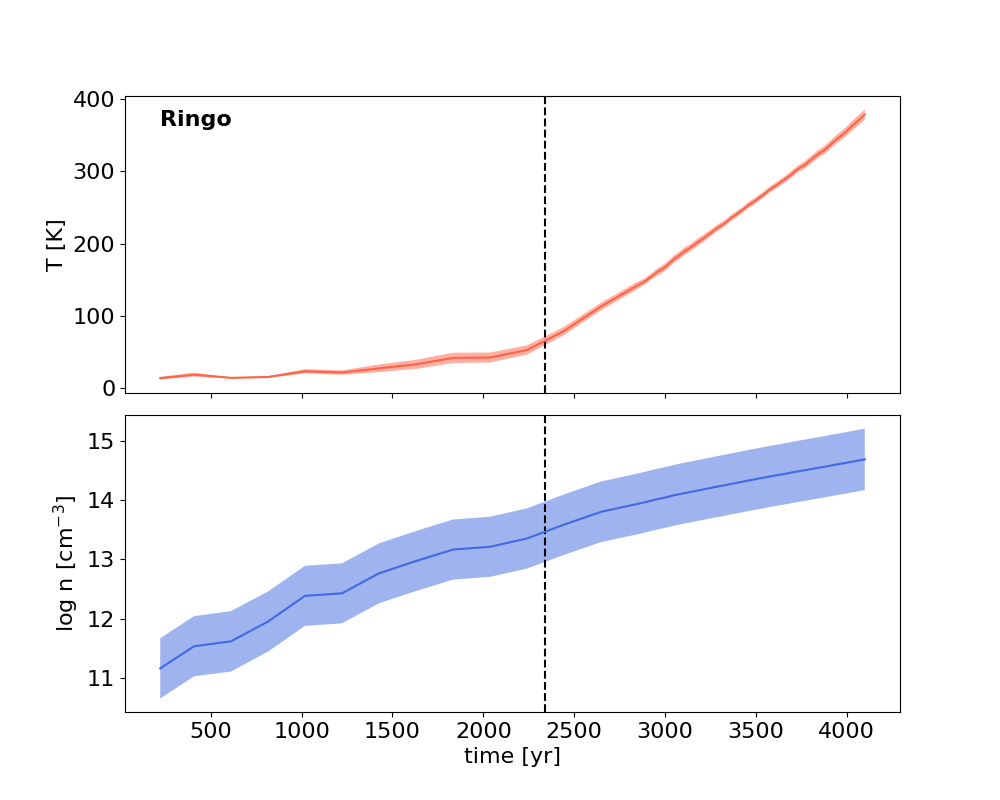}

\caption{Temperature (top panels) and number density (bottom panels) of the four fragments calculated from their constituent particles (see Table \ref{tab:fragment_particles}).  Mean values are shown with a solid line, while the shaded region indicates the standard deviation.  The dashed vertical lines indicate the formation, and where appropriate, the destruction times for each fragment. All fragments experience increasing temperatures and densities during formation, but the material that constituted John and Paul cools and rarefies rapidly after tidal disruption at approximately 3000 and 3500 yrs, respectively.}
\label{fig:fragment_phys}
\end{figure*}

\subsection{Chemistry}

In order to calculate the chemistry of the hydrodynamic model, we adopt the approach developed in \citet{ilee_2011}, including the improvements implemented in \citet{evans_2015}, which we also describe here.  The chemical network consists of 125 chemical species and 1334 reactions.  The reactions were originally selected from the UMIST95 database \citep{millar_1997}, but their rate co-efficients were updated using data from the Kinetic Database for Astrochemistry \citep[KIDA\footnote{http://kida.obs.u-bordeaux1.fr/},][]{wakelam_2012}.  Various gas phase chemical processes are considered, including positive ion-neutral reactions, ionisation by cosmic rays, charge transfer, proton transfer, hydrogen abstraction, radiative and dissociative recombination with electrons, radiative association, neutral exchange, photodissociation and photoionisation.  

\smallskip

Due to the unknown UV properties and embedded nature of young Class 0 sources, we assume that the disc is well shielded from sources of external irradiation (setting $A_{\mathrm{V}}$ = 100) but allow photochemical reactions induced by cosmic rays, the ionisation rate of which is assumed to be $\zeta = 10^{-17}$~s$^{-1}$ throughout the disc.  In addition, several gas-grain processes are implemented, including adsorption, desorption, recombination of ions with negative grains, and very simple grain surface chemistry (e.g.\ the rapid hydrogenation of adsorbed species).  The chemical model assumes a constant gas-to-dust ratio of 100, and a uniform distribution of grains with radius 0.1\,$\mu$m. 

\smallskip

While grain surface chemistry likely plays an important role in the evolution of many chemical species, the chemical pathways are uncertain and computationally expensive to implement.  In order to simulate any further potential effects of grain surface reactions, our initial gas phase species abundances are taken to be representative of observations of cometary ice abundances \citep[see][]{ehrenfreund_2000}, given in Table \ref{tab:initial_abundances}.  While there is some consistency between cometary and interstellar ices, it is not known whether comets undergo significant chemical processing, and thus, whether they accurately preserve the initial elemental compositions of young systems is not clear \citep[see, e.g., ][]{caselli_2012}.

\smallskip

\begin{table}
    \centering
    \begin{minipage}{0.65\columnwidth}
        \centering
        \caption{Initial gas phase fractional abundances $X(i) = n_{i} / n$ and molecular binding (desorption) energies, E$_{\rm b}$.  Species omitted here were initialised with a fractional abundance of $10^{-20}$.}
        \begin{tabular}{ccc}
        \hline
        Species     &   $X(i)$                  & E$_{\rm b}$ (K)           \\
        \hline
        He          &   $1.00\times 10^{-1}$    & $-$                   \\
        CO          &   $3.66\times 10^{-5}$    & 855                   \\
        CO$_{2}$    &   $3.67\times 10^{-5}$    & 2575                  \\
        H$_{2}$O    &   $1.83\times 10^{-4}$    & 5773                  \\
        H$_{2}$CO   &   $1.83\times 10^{-6}$    & 2050                  \\
        CH$_{4}$    &   $1.10\times 10^{-6}$    & 1100                  \\
        HCN         &   $4.59\times 10^{-7}$    & 2050                  \\
        HNC         &   $7.34\times 10^{-8}$    & 2050                  \\
        S           &   $1.62\times 10^{-5}$    & 1100                  \\
        NH$_{3}$    &   $3.30\times 10^{-6}$    & 5534                  \\
        H$_{2}$S    &   $2.75\times 10^{-6}$    & 2743                  \\  
        SO          &   $1.47\times 10^{-6}$    & 2600                  \\
        SO$_{2}$    &   $1.84\times 10^{-7}$    & 3405                  \\
        OCS         &   $3.30\times 10^{-6}$    & 2888                  \\
        \hline
        \end{tabular}
        \label{tab:initial_abundances}
    \end{minipage}    
\end{table}

All SPH particles from the full hydrodynamic simulation ($4 \times 10^{6}$ in total) are used as physical inputs for the chemical model, providing temperatures and densities as a function of space and time.  The chemical model interpolates these quantities (and other parameters relevant for the full output of the physical model) on to timescales which are more appropriate for the chemistry.  This approach ensures that the chemical evolution timesteps are not fixed to the timesteps used in the physical model, allowing very rapid chemical processes to be followed with sufficient temporal resolution.

\smallskip

The DVODE integrator \citep{brown_1989} is used to integrate the chemical rate equations, yielding fractional abundances for each species, e.g. $X_i$ = $n_i$/$n$, where $n$ = $n_{\mathrm{H}} + n_{\mathrm{He}} + n_{\mathrm{Z}}$.  In order to improve the success of the integrations, we adopt the recursive timestep algorithm implemented in \citet{evans_2015}.  Integrations which are initially unsuccessful are re-attempted by continually halving the timestep until either the integration is successful, or the timestep becomes so small that integration is not possible.  Using this approach, we were able to achieve very high success rate, with less than 0.3 per cent of tracer particles failing to complete integration.  These failed particles included very rapid jumps in temperature or density, but were evenly distributed across the simulation and so their exclusion had a negligible effect on the final chemical abundances throughout the disc.  In order to visualise the results of the chemical modelling of the particles, we employ a mapping and interpolation method similar to that of \citet{price_2007} to display abundances and column densities.  

\section{Results}
\label{sec:results}

\subsection{Global evolution of the disc}

Figure \ref{fig:hydro_evolution} displays several snapshots of the disc evolution.  After approximately 1300 years of evolution, corresponding to approximately 1.5 outer rotation periods (ORPs), the disc fragments into a large number of objects, with masses of order 0.5 -- 4~$\mjup$.  Most of these objects are transient, being quickly sheared out by the disc.  Of the eight or so seen in the upper left panel of Figure \ref{fig:hydro_evolution}, only four remain on relatively long timescales, which we identify using clump finding algorithms \citep{hall_2017}.  We label them (in ascending distance from the protostar) --- John, Paul, George and Ringo --- in the middle left panel\footnote{Our simulation is identical to that of `Simulation 2' described in \citet{hall_2017}, in which the fragments are named Clump 5, 3, 4 and 2 in increasing radial distance from the central star, respectively.}.  

\smallskip

Of these four, John and Paul migrate inwards, and are tidally disrupted at $t \sim 3100$~yr and $t \sim 3500$~yr respectively.  George and Ringo survive until the end of the simulation run at $t=4100$~yr.  George migrates inwards and accretes a significant amount of inner disc material, finally approaching the initial radial distance attained by Paul when the four fragments are first identified.  Ringo orbits in relatively low density surroundings, and as such only accretes material within a few au.  Ringo ends the simulation at a greater distance than where it began, due to dynamical interactions with George and the intervening spiral structure.

\smallskip

It is important to note that the fate of George is not assured here.  While this fragment has had considerably longer to become fully gravitationally bound (compared to John and Paul), it may well be the case that it still migrates inwards towards tidal disruption, perhaps achieving a closer distance to the star before being destroyed.  On the other hand, Ringo appears to be safely ensconced at large distances, and it may be joined by a fifth fragment in the beginnings of formation at $\sim 87$~au when the simulation is ended (see final panel of Figure \ref{fig:hydro_evolution}).  Ideally we would continue the simulation further to track the fifth fragment, but the densities reached by the other fragments would require us to allow sink particle creation, which would obstruct our efforts to track both fragment structure and chemistry.

\smallskip

Figure \ref{fig:temperature} displays the disc mid-plane temperature at $t=2852$~yr and $t=4100$~yr, corresponding to the third and final panels of Figure \ref{fig:hydro_evolution}.  Globally, the disc retains the initial radially-decreasing temperature profile which is expected from less massive discs.  However, significant deviations from this profile are seen toward the spiral arms and the fragments, where extra heating is caused by compressive forces and weak shocks.

\subsection{Physical conditions within the fragments}
\label{sec:phys_fragments}

We identify the population of particles that constitutes the four fragments by considering a simulation snapshot taken at $t=2852$~yr, and collecting all particles that reside within 1~au of the central density peak.  The initial locations of the density peak and total particle counts can be found in Table \ref{tab:fragment_particles}.  The fragment locations are consistent with the results derived from more advanced clump finding algorithms that have been used on this dataset in other work \citep[see][]{hall_2017}, which we use to report the fragment masses in Table \ref{tab:fragment_particles}.  Our approach is likely to be an undersampling of the material bound to each fragment -- the typical Hill radii of these bodies is around 5~au or larger, and 95 per cent of the mass of George and Ringo is contained within approximately 2.5\,au at the end of the simulation.  However, this approach ensures that all material for which we study the chemical composition is certain to remain inside the fragment after formation, and as such will not include any material that is only incorporated into a fragment for a short time.

\smallskip

\begin{table}
    \centering
    \begin{minipage}{\columnwidth}
    \caption{The position of the fragments at at $t = 2851$\,yrs, along with the number of particles $N$ within a cylindrical radius $<$ 1\,au of the central density peak at 2852\,yr.  These particles are subsequently used to sample the physical and chemical conditions of the fragments.  Also given are the total masses of the fragments $M$ at $t = 2852$\,yrs calculated from \citet{hall_2017}, and the formation ($t_{\rm f}$) and destruction ($t_{\rm d}$) times for each fragment.}
    \label{tab:fragment_particles}
    \begin{tabular}{lccccccc}
         \hline
           & \multicolumn{3}{c}{Position (au)}                &      $N$                        & $M$               & $t_{\rm f}$   & $t_{\rm d}$       \\
                  & \white{$+$}$x$ & \white{$+$}$y$ & \white{$+$}$z$ &                          & (M$_{\rm Jup}$)   &  (yr)         &  (yr)     \\
         \hline
         John     & \white{$+$}$42.9$      & \white{$+$}$10.3$        & $-0.2$   & 27033            & 10.3          & 1935          & 3106      \\   
         Paul     & \white{$+$}$22.3$      & $-55.0$                  & \white{$+$}$0.0$    & 13424 & \phantom{0}3.7           & 1528          & 3514      \\ 
         George   & \white{$+$}$90.2$      & \white{$+$}$40.6$        & $-0.1$   & 29603            & \phantom{0}8.2           & 1935          & --        \\   
         Ringo    & $-99.2$                & $-16.4$                  & \white{$+$}$0.1$   & 16902  & \phantom{0}5.4           & 2342          & --        \\        
         \hline
    \end{tabular}
    \end{minipage}
\end{table}

Figure \ref{fig:fragment_phys} shows the averaged temperature and number density of each fragment, calculated from the collections of particles outlined in Table \ref{tab:fragment_particles}.  During formation, all fragments experience increasing temperatures and densities as they contract.  These temperature and density enhancements above the mean disc value persist throughout their lifetime. As John and Paul are disrupted by tidal shearing at approximately 3000 and 3500 years respectively, their constituent material cools and rarefies rapidly.  Following disruption, this material is distributed throughout the inner disc and experiences, on average, relatively steady temperature and density conditions.  The fragments further out in the disc, George and Ringo, continue to become hotter and more dense as they orbit in the disc.  In particular, George continues to accrete significant material from the disc, with a mass in excess of 25 $\mjup$ by the end of the simulation.  

\smallskip

In order to examine the physical conditions as a function of radius within the fragments, we performed a spherical average over the particles that comprise each fragment for each timestep during the simulation.  The results of this averaging are shown in Appendix \ref{app:frag}, Figure \ref{fig:fragphys}.  At the time of classification ($t=2852$\,yrs) all fragments exhibit a radially decreasing temperature and density profile.  Paul and Ringo show an upturn in temperature profile at distances $\sim$ 1\,au from the centre.  This is caused by the increasingly large proportion of particles at higher $z$, and thus higher temperature, than the remaining small fraction of material at these radii.  Once temperatures in the centre of the fragments rises sufficiently, this effect becomes insignificant.  At the end of the simulation, George has contracted significantly to $\sim$ 0.5\,au, reaching a central temperature of 1100\,K and number density of $3\times10^{15}$\,cm$^{-2}$ which sharply decline with increasing distance from the fragment centre. In contrast, Ringo has not contracted significantly, and instead shows a somewhat flatter temperature profile, with $\sim$ 400\,K being sustained out to approximately 0.5\,au before beginning to decrease.

\subsection{Chemical evolution of the disc and fragments}
\label{sec:chem_disc}

\begin{figure*}
\includegraphics[height=0.38\textwidth, trim={1.05cm 0.4cm 1cm 0.8cm},  clip]{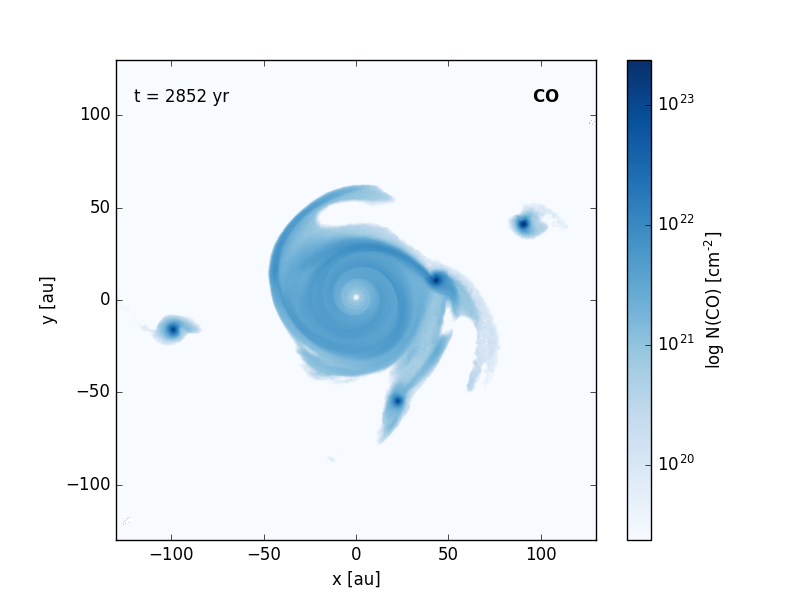}
\includegraphics[height=0.38\textwidth, trim={1.05cm 0.4cm 1cm 0.8cm},  clip]{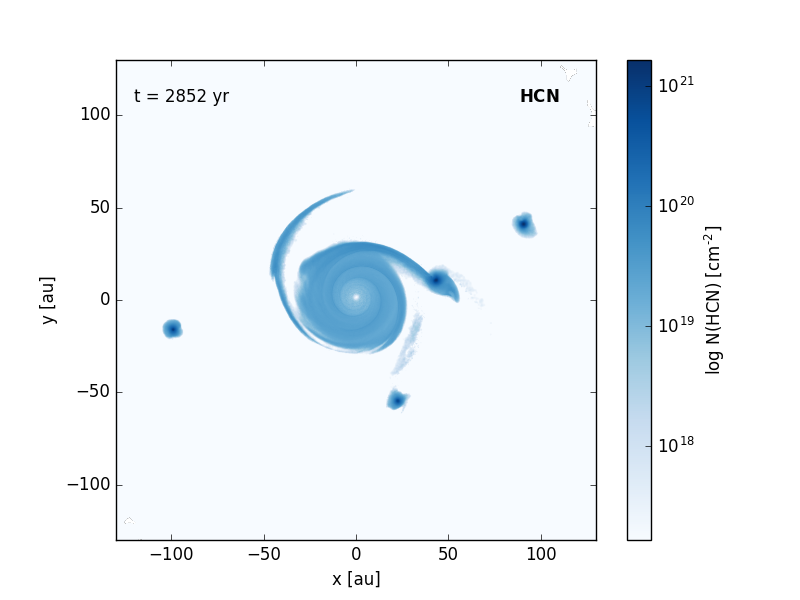} \\
\includegraphics[height=0.38\textwidth, trim={1.05cm 0.4cm 1cm 0.8cm},  clip]{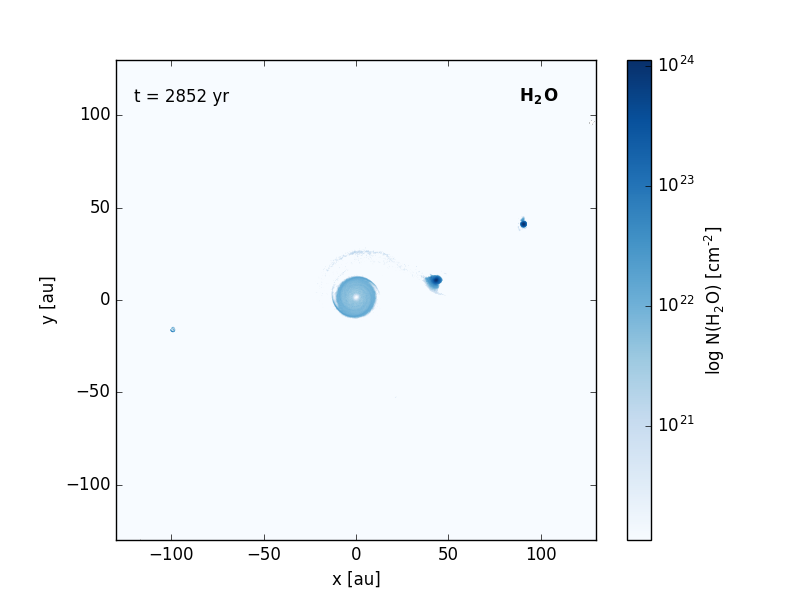}
\includegraphics[height=0.38\textwidth, trim={1.05cm 0.4cm 1cm 0.8cm},  clip]{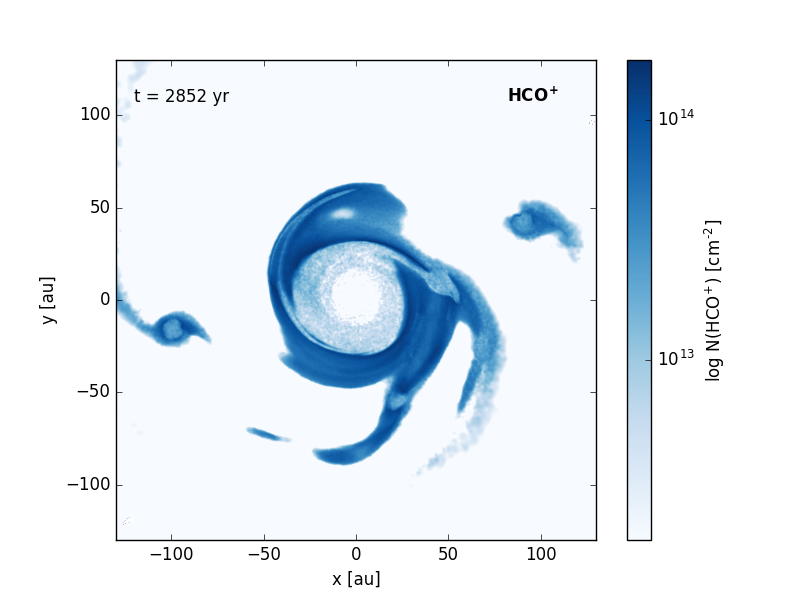} \\
\caption{Column density of selected chemical species (CO, HCN, H$_2$O and HCO$^{+}$) in the disc at t = 2851 years.  See Appendix \ref{app:chem} for the remaining species.}
\label{fig:selected_chem}
\end{figure*}

Figure \ref{fig:selected_chem} shows the column density for selected chemical species that we have modelled (plots for all species in our network can be found in Appendix \ref{app:chem}).  From examination of the temporal evolution of our simulation, we split the chemical species into morphological categories -- those which are primarily abundant towards the fragments (e.g. H$_{2}$O, H$_{2}$S, HNO, N$_{2}$, NH$_{3}$, OCS, SO) and those which are also abundant in the spiral shocks within the disc (e.g. CO, CH$_{4}$, CN, CS, H$_{2}$CO).  Based on representative particles that reside within 1 au of the fragments, we can examine the creation and destruction routes of various species.  Following the chemical evolution of an SPH particle that becomes incorporated within George at $t=2750$ years, we can see that CN is primarily formed via secondary photodissociation of HCN and HNC, and destroyed by reactions with NH$_{3}$ forming NH$_{2}$ and HCN.  H$_2$O has many formation routes, but is most significantly formed by the reaction of H$_{3}$O$^{+}$ and NH$_{3}$ and removed from the gas phase via adsorption.  HCO$^{+}$ is primarily created by reactions of H$_{3}^{+}$ with CO.  It is readily destroyed when H$_{2}$O exists in significant quantities in the gas phase, forming H$_{3}$O$^{+}$ and CO.  Adsorption and desorption primarily determine the gas phase abundances of CO, CO$_{2}$, CS, HCN, H$_{2}$CO, HNO, NH$_{3}$, OCN and OCS.  The timescales for the adsorption and desorption range from a few seconds to fractions of a year, depending on the temperature and density of the ambient material.  As in \citet{ilee_2011} and \citet{evans_2015}, this is many orders of magnitude smaller than the local dynamical timescales within the disc, where an outer rotation period takes approximately 1000 years.

\smallskip

Within the regions of the disc that have not undergone fragmentation, we identify two broad categories of molecules.  Firstly, there are those molecules in which the fractional abundance is primarily determined by adsorption and desorption processes (e.g.\ CO, CO$_{2}$, SO, SO$_{2}$, NH$_{3}$, H$_{2}$O, H$_{2}$CO, H$_{2}$S, HCN, HNC and OCS).  The abundances of these molecules are therefore almost entirely set by temperature changes within the disc.  Secondly, there are molecules which are significantly affected by gas-phase reactions occurring in warm, shocked regions (e.g.\ CS, HCS, HCS$^{+}$, HCO$^{+}$, HNO, OCN, O$_{2}$ and CN).  The higher temperatures experienced as the shocks progress through the disc allow some reactions with high activation energies to proceed at significantly enhanced rates.  Examples of molecules produced by these effects include --- OCN which is formed by a reaction of CO and N$_{2}$; HNO which is formed by NH$_{2}$ and O; and HCS which is formed by reactions of S and CH$_{2}$.  We note that our choice of initial gas-phase abundances (see Table \ref{tab:initial_abundances}) likely sets the molecules whose abundance is primarily determined by adsorption and desorption, and changing these initial abundances would impact the molecules that reside within this category.

\smallskip

Using the collections of particles from Table \ref{tab:fragment_particles}, we are able to follow the average chemical evolution of the material that constitutes each fragment - before, during and (where appropriate) after the fragment has been disrupted.  Figure \ref{fig:av_frag_chem} shows the average fractional abundance of various species (CO, H$_{2}$O, CN, HCO$^{+}$, HCS) for each of the fragments.  Because of the rapid temperature increase in almost all of the material that becomes a fragment, CO is quickly liberated from ice mantles and remains in the gas phase.  H$_{2}$O exhibits very similar evolution within John, George and Ringo -- mostly located on the grains even as the fragments are fully formed initially, but then quickly liberated by thermal desorption as the fragment temperatures increase.  The exception to this behaviour is Paul, which does not reach sufficiently high temperatures for the H$_{2}$O abundance to reach peak values before undergoing tidal disruption.  After disruption in John and Paul, the H$_{2}$O abundance quickly drops again as the molecule freezes out.  The abundance of CN is initially high in all fragments, but this quickly drops when temperatures rise above $\sim$150\,K due to reactions with thermally desorbed NH$_{3}$ creating HCN and NH$_{2}$.  The HCO$^{+}$ abundance in all fragments is low due to reactions with the ever increasing amounts of H$_{2}$O in the gas phase, but is boosted as John and Paul undergo disruption as material cools and the H$_{2}$O begins to freeze out.  The abundance of HCS remains relatively constant throughout the lifetimes of the fragments, with only a small drop during the disruption for John and Paul.

\smallskip

The spherical averaging discussed in Section \ref{sec:phys_fragments} also allows us to examine radial variations of the selected species within each fragment, shown in Appendix \ref{app:frag}, Figure \ref{fig:fragchem}.  At the time of classification ($t=2852$\,yrs), John and George exhibit very small radial differences in the fractional abundance of CO, H$_{2}$O, CN, HCO$^{+}$ and HCS due to their relatively high temperatures.  Paul and Ringo exhibit a more varied chemical structure, in particular showing a lower peak abundance of H$_{2}$O, that then decreases with increasing distance from the fragment centre.  Paul also shows an increasing abundance of CN at larger radii.  For the fragments that survive to the end of the simulation, any initial radial variations in these species seem to have mostly been eradicated by the increasing temperatures and densities.      

\smallskip

It is important to note that the temperatures within George rise outside the range of temperatures for which our chemical network is likely valid.  Determining chemical abundances for these temperatures ($T > 300$--$500$\,K) requires dedicated high temperature reactions, which are not included within our network, and as such the specific abundances reported for George after approximately 3000\,yr should be treated with caution.

\begin{figure*}
\includegraphics[height=0.3\textwidth, trim={0.5cm 0cm 4cm 1.0cm},  clip]{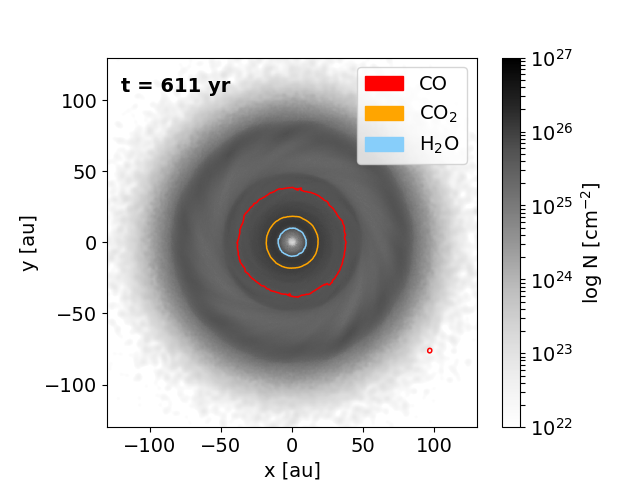}
\includegraphics[height=0.3\textwidth, trim={2.65cm 0cm 4cm 1.0cm},  clip]{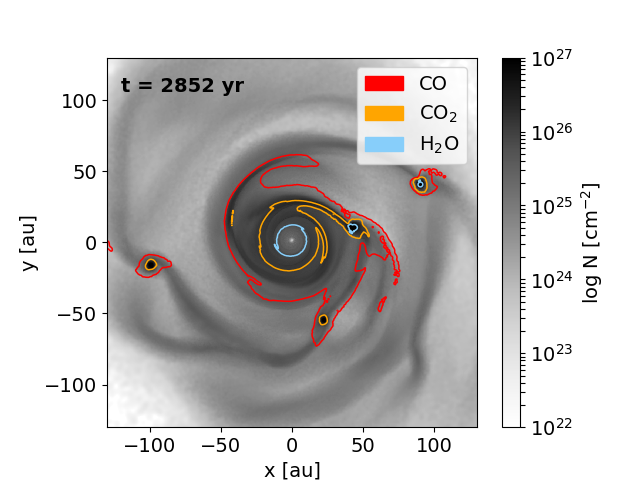}
\includegraphics[height=0.3\textwidth, trim={2.65cm 0cm 4cm 1.0cm},  clip]{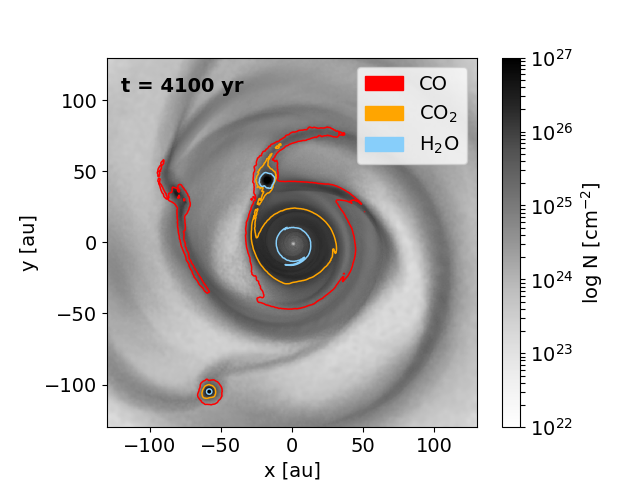}
\includegraphics[height=0.3\textwidth, trim={12.3cm 0cm 0.5cm 1.0cm}, clip]{figures/snowline_063.png}
\caption{Evolution of the total column density of the disc (grey scale, as in Figure \ref{fig:hydro_evolution}) overlaid with the `snow lines' of CO (red), CO$_{2}$ (orange) and H$_{2}$O (blue).  Before onset of the global instabilities, the disc shows the expected concentric ringed structure of snow lines.  However, significant deviations from this are seen once the spiral shocks set in and fragmentation begins.  In particular, the fragments develop concentric snow lines themselves.}
\label{fig:chemical_evolution_snowline}
\end{figure*}

\section{Discussion}
\label{sec:discussion}

\subsection{Comparison with non-fragmenting discs}

Direct comparison of our model with other, non-fragmenting discs which are subject to gravitational instabilities is challenging due to the differing assumptions between the chemical networks, reaction rates and underlying hydrodynamic models used.  Nevertheless, here we outline some broad comparisons with previous work that characterises the chemical evolution of such discs.

\smallskip

Perhaps the closest simulation in terms of a disc-to-star mass ratio is that of \citet{evans_2015} in which $q \sim 0.2$ (compared to our $q = 0.25$).  This disc is much smaller in radial extent than the disc studied here, and therefore does not undergo fragmentation during the 2000\,yrs for which it is followed due to the higher ratio of cooling time to dynamical time at smaller radii.  The chemical network used in \citet{evans_2015} is identical to the one used in this work, and therefore comparisons made between their results and the results presented here are free from any differing assumptions on the nature of the chemistry.  When comparing CO, HCN, H$_{2}$O and HCO$^{+}$ column density maps from Figure \ref{fig:selected_chem} with the results from \citet{evans_2015}, it is clear that the general locations of maxima and minima are very similar if the fragments are disregarded.  In particular, the CO appears to trace the spiral arms well, the HCN slightly less so, the H$_{2}$O in confined to the hottest regions and the HCO$^{+}$ is most abundant in the outer regions of both discs.  Interestingly, the extent of the CO emission in our fragmenting disc model and in the model of \citet{evans_2015} is similar.  This suggests that the apparent size of discs in molecular line emission may be similar even for discs which initially possess very different radial extents, due to the location of the molecular condensation front (or snow line) which is tied to the thermal structure of the disc.  \citet{evans_2015} also note that the persistent shock heating can lead to permanent changes in the chemical composition of inner disc material.  Species affected include HNO, CN and NH$_{3}$. These effects are also likely to occur in our disc, but we are unable to characterise them due to the lower temporal resolution over which we record the physical information of our hydrodynamic model.  Such a lower resolution was essential in order to be able to evaluate the chemistry of the $4\times 10^{6}$ particles that comprise our disc simulation.  

\smallskip

\citet{hincelin_2013} studied the chemical evolution of a collapsing cloud that forms an gravitationally unstable disc around the first Larson core using radiation-magneto-hydrodynamics and a gas-grain chemical network.  While column density plots for their model are not presented, the fractional abundances of the CO and H$_{2}$O ices are shown.  Contrary to our results, the disc in \citet{hincelin_2013} does not seem to reach the temperatures required in the spiral arms to desorb significant amounts of either molecule into the gas phase, leading to large abundances of their ices in the spiral arms.   

\subsection{Molecular condensation fronts (a.k.a. `snow lines')}

\begin{figure*}
\includegraphics[width=\columnwidth, trim={0.45cm 0.25cm 0.75cm 0.75cm},  clip]{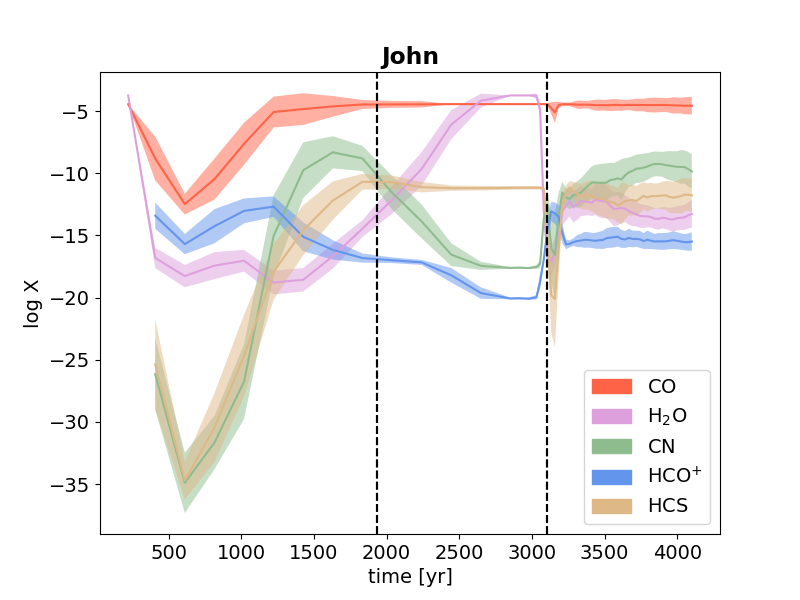}
\includegraphics[width=\columnwidth, trim={0.45cm 0.25cm 0.75cm 0.75cm},  clip]{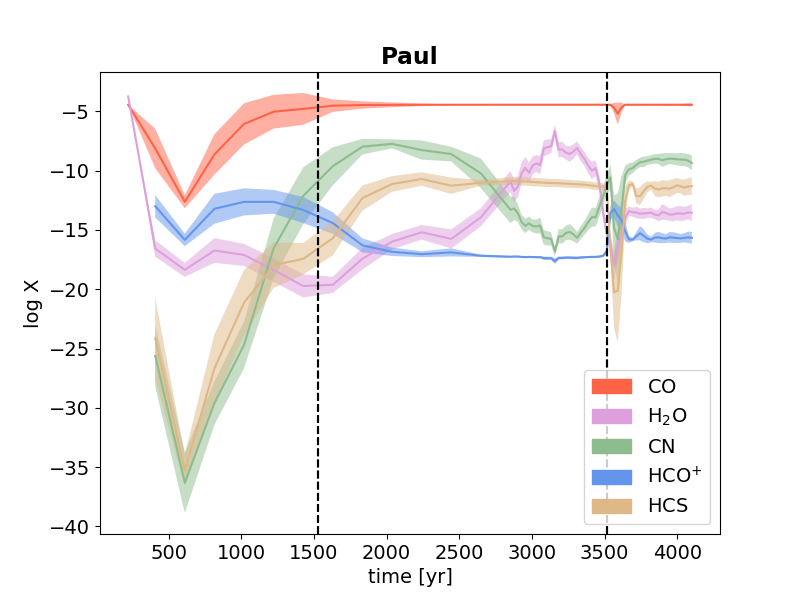}

\includegraphics[width=\columnwidth, trim={0.45cm 0.25cm 0.75cm 0.75cm},  clip]{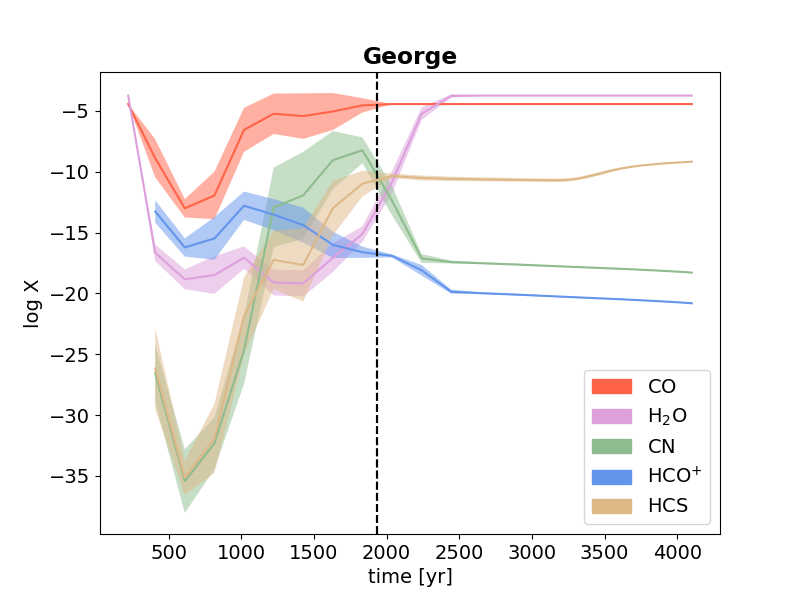}
\includegraphics[width=\columnwidth, trim={0.45cm 0.25cm 0.75cm 0.75cm},  clip]{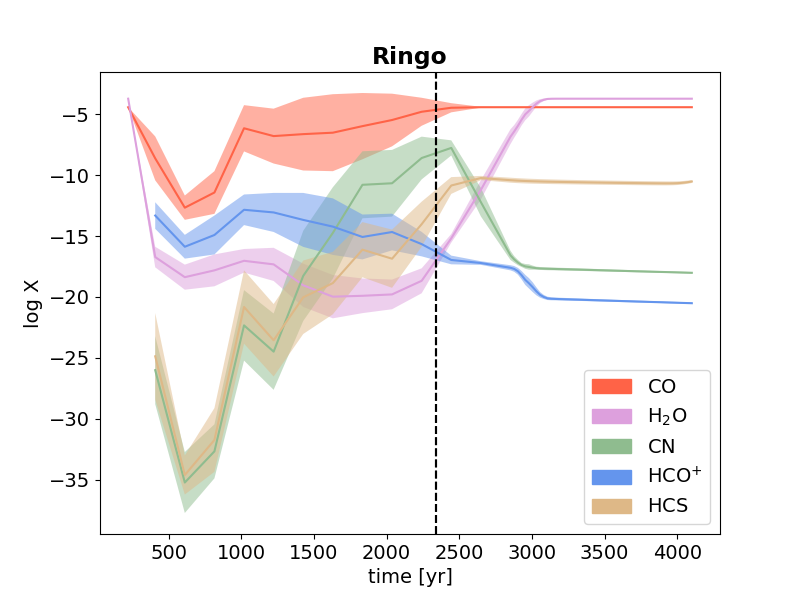}
\caption{Average fractional abundance of selected species contained within each fragment.  Mean values are shown with a solid line, while shaded regions indicate the standard deviation.  The dashed vertical lines indicates the formation and, where applicable, destruction times for each fragment.}
\label{fig:av_frag_chem}
\end{figure*}

The locations of the condensation fronts at which molecular species transition from the gas-phase to ices on grain surfaces, so-called `snow lines', are important because the properties of the dust grains on either side may be drastically different \citep[see, e.g.,][]{qi_2013, banzatti_2015, eistrup_2016, cieza_2016, panic_2017}.  Such differences lead to changes in the raw ingredients available for planet formation via core accretion and associated processes which involve the agglomeration of small particles \citep[e.g.][]{oberg_2011, madhusudhan_2012,booth_2017}.   In quiescent discs, these locations are often quoted as a single radial location corresponding to where the mid-plane temperature drops below the temperature at which species begin to rapidly freeze-out (see e.g.\ \citealt{khajenabi_2017}).  However, in discs which undergo significant non-axisymmetric temperature and density increases, these snow lines may appear in multiple locations, generally following the regions at which the temperature corresponds to the freeze-out temperature for the molecule of interest.  In particular, we note that in our model, these locations trace a three dimensional structure across the $x$, $y$ and $z$ directions.  Therefore, in reality, a more appropriate term to describe the phenomenon in such discs might be a snow \emph{`bubble'}.  However, due to the relatively low vertical resolution in our disc, we restrict our description of the condensation fronts to the radii at which the gas-phase column density of the species in question is equal to that of the column density of the corresponding ice-phase species, i.e.\ $N_{\rm gas}$ = $N_{\rm ice}$.

\smallskip

Figure \ref{fig:chemical_evolution_snowline} shows the evolution of the column density of the disc overlaid with contours showing the location of the mid-plane `snow lines' for CO, CO$_{2}$ and H$_{2}$O.  Early in the simulation, before the onset of global instabilities, the disc experiences the expected pattern of concentric circular snow lines.  The H$_{2}$O snow line is closest to the central star, followed by the CO$_{2}$ and finally the CO snow lines -- the ordering determined by their decreasing binding energy to icy grain surfaces (see Table \ref{tab:initial_abundances} and \citealt{evans_2015}, their Table A1).  However, once the instabilities develop, the temperature structure of the disc is altered significantly by shock heating, the formation of fragments, and compression under gravity.  The highest temperatures are seen toward the fragments and in the spiral arms.  In these regions, significant increases in the gas-phase abundances of molecules can be seen (e.g. Figure \ref{fig:chemical_evolution_snowline}, right panel).  These regions would, without the instabilities, lie in a temperature and density regime in which these molecules would be frozen out.   

\smallskip

Observationally, this would manifest itself as an apparent increase in the size of the emitting region of these molecules.  While in this particular disc the radial increase is relatively small (moving from approximately 40 to 60\,au), discs which do not undergo fragmentation can exhibit increases that are much larger \citep[e.g.][]{ilee_2011, evans_2017}.  Indeed, such increases in gas-phase abundances of volatile molecules caused by dynamic effects may have already been seen toward young embedded discs in the Perseus molecular cloud \citep{frimann_2017}.  Similar effects have been seen in other models \citep[e.g.][]{Cleeves_2015}, however these have considered the temperature changes induced in the disc by the radiative output of an already-formed planet, rather than the dynamic effects occurring during the formation stages which we model here.

\subsection{The C/O ratio of the dust and gas components}
\label{sec:co_ratio}

\begin{figure*}
\includegraphics[width=\columnwidth, trim={0.75cm 0.25cm 0.75cm 0.75cm},  clip]{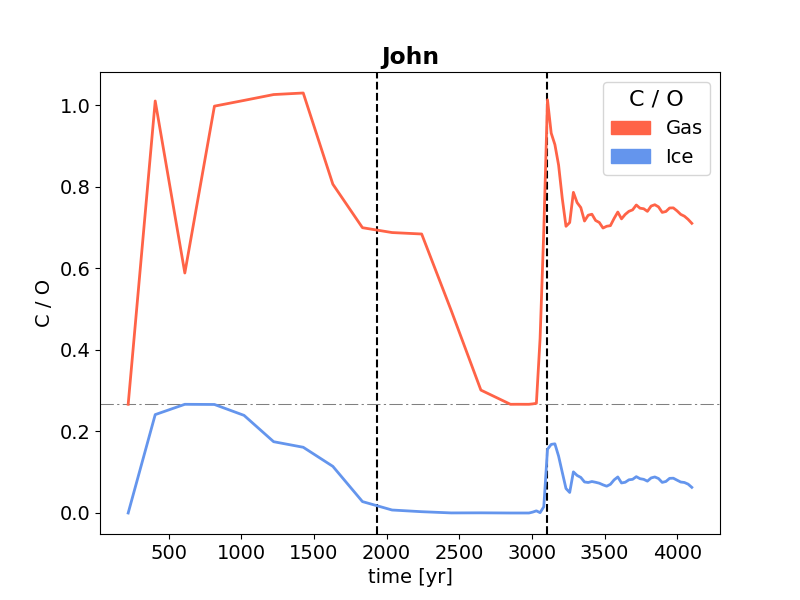}
\includegraphics[width=\columnwidth, trim={0.75cm 0.25cm 0.75cm 0.75cm},  clip]{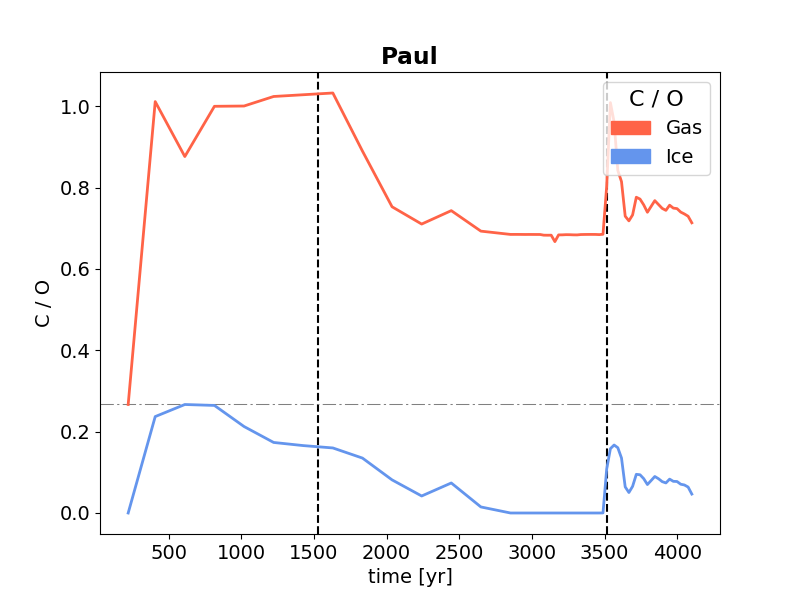}

\vspace{1em}

\includegraphics[width=\columnwidth, trim={0.75cm 0.25cm 0.75cm 0.75cm},  clip]{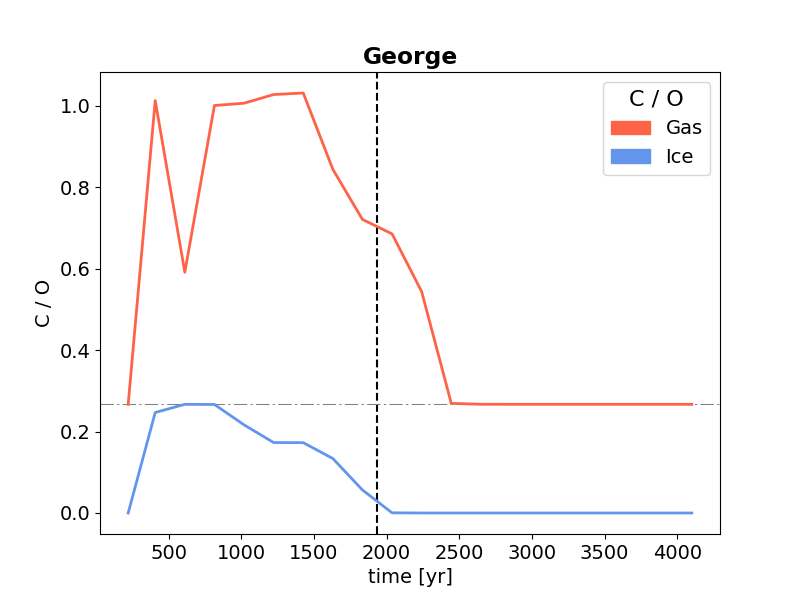}
\includegraphics[width=\columnwidth, trim={0.75cm 0.25cm 0.75cm 0.75cm},  clip]{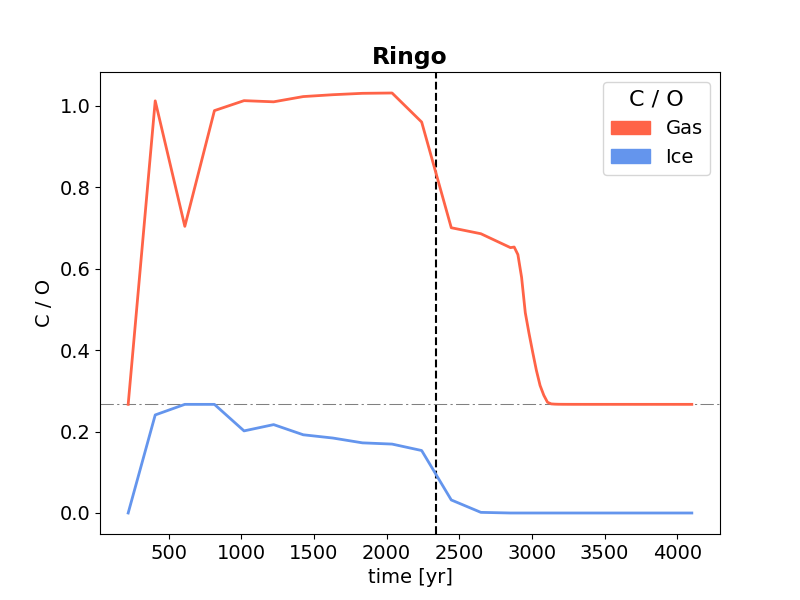}
\caption{Average C/O ratio for each fragment (John, Paul, George, Ringo, clockwise from top left) including species in the gas phase (red) and on the grain surfaces (blue).  The dashed vertical lines indicate the formation and, where appropriate, the destruction times for each fragment, and the horizontal dash-dot line indicates the initial C/O ratio of 0.26.}
\label{fig:av_CO_ratio}
\end{figure*}

\smallskip

The total ratio of carbon to oxygen (C/O) can be an important quantity in characterising the properties of a planetary atmosphere, in particular where in a protoplanetary disc the planet may have formed \citep{oberg_2011, madhusudhan_2012, booth_2017}.  Using the collections of particles in Table \ref{tab:fragment_particles}, we are able to calculate the bulk average [C/O] for each of the fragments as a function of time for both the gas and ice phase (see Figure \ref{fig:av_CO_ratio}).  Globally, the general [C/O] ratio evolution can be understood as follows.  With our initial chemical abundances, the initial ratio of [C/O]$_{\rm gas}$ is 0.26\footnote{We note that our choice of initial chemical conditions (see Table \ref{tab:initial_abundances}) influences our initial elemental ratios of the disc material, in particular setting our initial C/O ratio to 0.26.  This value is lower than measurements of the solar C/O ratio, determined to be closer to 0.5 \citep{prieto_2002}.  We stress that changing this initial value would not alter the qualitative outcomes presented here, and would simply vertically shift the baselines plotted in Figure \ref{fig:av_CO_ratio}.} and undefined for [C/O]$_{\rm ice}$ due to all species initially being in the gas phase.  Particles almost immediately diverge from from this initial value, with [C/O]$_{\rm gas}$ rising to unity, as CO remains in the gas phase while other species with higher binding energies freeze out very efficiently onto the surfaces of dust grains (raising the [C/O]$_{\rm ice}$ to almost the initial gaseous value).   As temperatures begin to increase, CO$_{2}$ is desorbed from the grain surfaces leading to a drop in [C/O]$_{\rm gas}$ at the formation stages of each of the fragments.  Further increases in temperature cause H$_{2}$O to desorb from the grain surfaces, returning the [C/O]$_{\rm gas}$ back to the initial value of 0.26, finally leaving little-to-no C or O on the grain surfaces in material that continues to experience high temperatures.  For material that experiences a drop in temperatures, the reverse process occurs.  Thus, the final composition of the material which goes on to make up a fragment eventually preserves the initial elemental ratio.     

\smallskip

However, because we are able to follow the temporal evolution of the [C/O] in our model, we observe interesting transient features.  During the early formation stages of each fragment, the average temperatures are not high enough to entirely desorb all species from the grain surfaces.  This `cold start' lasts for between 500--2000\,yrs, and as a consequence the [C/O]$_{\rm gas}$ is held at intermediate values between unity and the initial value of 0.26.  In fact, Paul does not exist within the disc for a long enough period of time to reach temperatures that ensure a return of [C/O]$_{\rm gas}$ to the initial value (due to H$_{2}$O remaining frozen out) before being tidally disrupted.
 
\smallskip

This effect may have important implications for the composition of any resulting objects that are formed.  It has been shown that it is possible to settle dust grains toward the centres of non-turbulent gaseous planetary cores \citep[e.g.][]{boss_1998, boley_2010, nayakshin_2010b}.  Therefore, the transient features we observe in [C/O]$_{\rm ice}$ are of extreme interest because, if the timescale for dust to settle is comparable to the timescale of the transient feature, then ices with non-primordial [C/O] ratios may be incorporated into these solid cores.  From \citet{nayakshin_2010b}, the timescale for sedimentation of dust, $t_{\rm sed}$, is given by 
\begin{equation}
t_{\rm sed} \approx 5 \times 10^{3} \dfrac{M_{\rm c}}{0.01 M_{\odot}} \left( \rho_{\rm d} \, a \right)^{-1} {\rm years}, 
\end{equation}
where $M_{\rm c}$ is the mass of the core, $\rho_{\rm d}$ is the density of the dust grains, and $a$ is the dust grain radius (where the latter two quantities are defined in cgs units).  Assuming $\rho_{\rm d} = 2.3\,{\rm g}\,{\rm {cm}^{-3}}$ which would imply silicate grains, the corresponding sedimentation timescale for centimetre-sized grains and larger would be 1000--2000 years for all fragments, and less than 500 years for 5\,cm grains or larger.  Such a timescale is comparable to the time over which H$_{2}$O can remain frozen out during the initial formation stages of the fragments (see Figure \ref{fig:av_frag_chem}).  In addition, grains which are covered in water ice mantles will have a higher sticking efficiency than bare grains \citep[e.g.][]{ehrenfreund_2003, wang_2005}.  Therefore, in the envelopes of fragments which exhibit this `cold-start', it may be possible to grow grains more quickly, increasing the fraction of solid material that can be sedimented toward the centre.  Thus, these fragments may be able to form larger rocky cores than might be expected, even if the eventual fate of the fragment as a whole is to be tidally disrupted.      

\smallskip

The C/O evolution discussed here not only has implications for the first few thousand years of evolution, but also for the final C/O ratio of the planet itself.  If dust can be efficiently settled to the fragment core before volatiles are evaporated from their surfaces, then changes in the C/O ratio can be expected -- higher values will be found in the envelopes, while lower values will be found in the core.  Unfortunately, the signature of this mechanism would be indistinguishable from what was discussed for the GI formation scenarios discussed in \citet{madhusudhan_2014}, in which different C/O ratios correspond to differing amounts of planetesimal accretion.  In all these cases, there is an inverse relation between the over-all metallicity of the planetary atmospheric gas (i.e.\ C/H and O/H) and the resulting C/O ratio.  As more ices are evaporated into the planetary atmosphere, the higher the metallicity, but the lower the C/O because the ices are rich in oxygen.  Only models that include differential transport of dust grains with respect to the gas can break this degeneracy \citep[see][]{booth_2017}.  We hope to extend our work with this treatment in the near future.   

\subsection{Chemical tracers of fragments}

Molecules with increased abundances toward fragments may offer a useful diagnostic of the physical conditions in these regions.  Such diagnostics will be important to probe in order to examine the planet formation process in detail.  \citet{narayanan_2006} have examined molecular line emission from a gravitationally unstable disc which includes a gas giant planet.  Using a hydrodynamic model coupled with a non-LTE radiative transfer code and fixed molecular abundances, they determine that the gas giant planet may be observable in high-$J$ transitions of HCO$^{+}$ if abundances are high enough.  However, in our model, we see significant depletion of HCO$^{+}$ due to the liberation of H$_{2}$O from grain surfaces by the high temperatures toward the fragments.  As such, in reality, observations of HCO$^{+}$ would likely not be an efficient tracer of these fragments, and if observed, would rather trace the circumfragmentary material around them instead.  

\smallskip

Recent spatially resolved observations of the disc around d216--0939 by \citet{factor_2017} have detected blue-shifted asymmetries in the HCO$^{+}$ and CO emission, which they suggest could be due to either a hydrodynamic vortex or the envelope of a forming protoplanet between 1.8--8\,M$_{\rm Jup}$ located at a radius of $60\pm20$\,au.  While \citet{factor_2017} point out that the disc of d216--0939 is likely to be currently gravitationally stable, the detected emission would be consistent with our results for the surviving fragments George and Ringo, and as such may be evidence that the disc in d216--0939 may have undergone fragmentation in the past. 

\smallskip

While we report here species that are increased in their fractional abundance toward the fragments, the assessment of their true observability requires molecular line radiative transfer modelling and synthetic imaging.  We will be performing a detailed study on the observable continuum and line emission in this model disc in a forthcoming publication (Hall et al., in preparation).

\section{Conclusions}
\label{sec:conclusions}

In this paper we have presented the first study of the chemical evolution in a fragmenting protoplanetary disc.  Using three dimensional hydrodynamics coupled with a non-equilibrium gas-grain chemical model, we have followed the chemical evolution of the $4\times 10^{6}$ particles that comprise the simulation.  We summarise our main findings as follows. 

\begin{itemize}

\item Our hydrodynamic model initially fragments into a large number of objects, of which four become quasi-stable.  The innermost two of these fragments migrate through the disc toward the central star and undergo tidal disruption, while the outermost two persist until the end of our simulation. 

\item The increased temperatures and densities found toward the spiral arms and fragments cause chemical reactions that would not otherwise take place at such large radii in the disc.  From these, we identify molecules that are abundant towards the fragments (e.g. H$_{2}$O, H$_{2}$S, HNO, N$_{2}$, NH$_{3}$, OCS, SO) and those which are also abundant in the spiral shocks (e.g. CO, CH$_{4}$, CN, CS, H$_{2}$CO).  In particular, HCO$^{+}$ may be a useful tracer of the circumfragmentary disc material.

\item Molecular condensation fronts, or `snow lines', deviate significantly from the expected concentric ring structures found in axisymmetric discs.  Increases in temperature causes by passing shocks desorb material at larger radii than would normally be expected, and fragments that have formed develop snow lines themselves.    

\item Fragments can exhibit a `cold start' after formation, with temperatures not high enough to entirely desorb H$_{2}$O from grain surfaces until after 500-2000\,yrs.  In one case, sufficiently high temperatures are not reached at all before the fragment in question undergoes tidal disruption.  In such cases, because the effect persists for a time comparable to the sedimentation time for dust in these fragments, grain growth in the envelopes of such fragments may be enhanced, leading to a larger proportion of solid material that can be sedimented toward the core.

\item This `cold start' phenomenon also gives rise to higher C/O ratios than would normally be expected for the envelopes of fragments formed via gravitational instability, and lower C/O ratios of any cores that may be formed toward the centre of the fragments.  As such, the atmospheric composition of planets formed via gravitational instability may not necessarily follow the bulk chemical composition of the disc from which they formed.    

\end{itemize}

Our results suggest that correctly accounting for the influence of the dynamical evolution on the corresponding chemical evolution of protoplanetary discs can provide useful information on the composition and properties of any resulting objects that are formed.  Such effects are important for the formation of giant planets and brown dwarfs, but also potentially for the formation of terrestrial planets by sedimentation within the fragments.  In the future, we plan to extend our modelling to account for variations in, for example, dust grain sizes and dynamics, allowing us to further probe the interlinked processes that occur as planets form via a variety of processes in young circumstellar discs.

\section*{Acknowledgements}
We would like to thank the referee for suggestions that improved the clarity of the manuscript, and Mihkel Kama for his reading of an early draft of this work.  JDI, RB and CJC gratefully acknowledge support from the DISCSIM project, grant agreement 341137, funded by the European Research Council under ERC-2013-ADG. DHF gratefully acknowledges support from the ECOGAL project, grant agreement 291227, funded by the European Research Council under ERC-2011-ADG.  DHF and WKMR also acknowledge support from STFC grant ST/J001422/1. MGE acknowledges a studentship funded by, and PC, TWH and AB acknowledge financial support from, the European Research Council (project PALs 320620).  CH gratefully acknowledges funding from the European Research Council (ERC) under the European Union's Horizon 2020 research and innovation programme (grant agreement No 681601).  Some of computational work for this paper was carried out on the joint STFC and SFC (SRIF) funded Wardlaw cluster at the University of St Andrews (Scotland, UK).  This research has  made  use  of   NASA's  Astrophysics  Data  System  Bibliographic Services and Astropy, a community-developed core Python package for Astronomy \citep{astropy_2013}.

\bibliographystyle{mnras}

\begin{thebibliography}{}
\makeatletter
\relax
\def\mn@urlcharsother{\let\do\@makeother \do\$\do\&\do\#\do\^\do\_\do\%\do\~}
\def\mn@doi{\begingroup\mn@urlcharsother \@ifnextchar [ {\mn@doi@}
  {\mn@doi@[]}}
\def\mn@doi@[#1]#2{\def\@tempa{#1}\ifx\@tempa\@empty \href
  {http://dx.doi.org/#2} {doi:#2}\else \href {http://dx.doi.org/#2} {#1}\fi
  \endgroup}
\def\mn@eprint#1#2{\mn@eprint@#1:#2::\@nil}
\def\mn@eprint@arXiv#1{\href {http://arxiv.org/abs/#1} {{\tt arXiv:#1}}}
\def\mn@eprint@dblp#1{\href {http://dblp.uni-trier.de/rec/bibtex/#1.xml}
  {dblp:#1}}
\def\mn@eprint@#1:#2:#3:#4\@nil{\def\@tempa {#1}\def\@tempb {#2}\def\@tempc
  {#3}\ifx \@tempc \@empty \let \@tempc \@tempb \let \@tempb \@tempa \fi \ifx
  \@tempb \@empty \def\@tempb {arXiv}\fi \@ifundefined
  {mn@eprint@\@tempb}{\@tempb:\@tempc}{\expandafter \expandafter \csname
  mn@eprint@\@tempb\endcsname \expandafter{\@tempc}}}

\bibitem[\protect\citeauthoryear{{Allende Prieto}, {Lambert}  \&
  {Asplund}}{{Allende Prieto} et~al.}{2002}]{prieto_2002}
{Allende Prieto} C.,  {Lambert} D.~L.,   {Asplund} M.,  2002, \mn@doi [\apjl]
  {10.1086/342095}, \href {http://adsabs.harvard.edu/abs/2002ApJ...573L.137A}
  {573, L137}

\bibitem[\protect\citeauthoryear{{Astropy Collaboration} et~al.,}{{Astropy
  Collaboration} et~al.}{2013}]{astropy_2013}
{Astropy Collaboration} et~al., 2013, \mn@doi [\aap]
  {10.1051/0004-6361/201322068}, \href
  {http://adsabs.harvard.edu/abs/2013A%26A...558A..33A} {558, A33}

\bibitem[\protect\citeauthoryear{{Banzatti}, {Pinilla}, {Ricci}, {Pontoppidan},
  {Birnstiel}  \& {Ciesla}}{{Banzatti} et~al.}{2015}]{banzatti_2015}
{Banzatti} A.,  {Pinilla} P.,  {Ricci} L.,  {Pontoppidan} K.~M.,  {Birnstiel}
  T.,   {Ciesla} F.,  2015, \mn@doi [\apjl] {10.1088/2041-8205/815/1/L15},
  \href {http://adsabs.harvard.edu/abs/2015ApJ...815L..15B} {815, L15}

\bibitem[\protect\citeauthoryear{Bate, Bonnell  \& Price}{Bate
  et~al.}{1995}]{bate_1995}
Bate M.~R.,  Bonnell I.~A.,   Price N.,  1995, MNRAS, 277, 362

\bibitem[\protect\citeauthoryear{{Bergin}, {Aikawa}, {Blake}  \& {van
  Dishoeck}}{{Bergin} et~al.}{2007}]{bergin_2007}
{Bergin} E.~A.,  {Aikawa} Y.,  {Blake} G.~A.,   {van Dishoeck} E.~F.,  2007,
  Protostars and Planets V, \href
  {http://adsabs.harvard.edu/abs/2007prpl.conf..751B} {pp 751--766}

\bibitem[\protect\citeauthoryear{Bodenheimer, Yorke, Rozyczka  \&
  Tohline}{Bodenheimer et~al.}{1990}]{bodenheimer_1990}
Bodenheimer P.,  Yorke H.~W.,  Rozyczka M.,   Tohline J.~E.,  1990, \mn@doi
  [ApJ] {10.1086/168798}, 355, 651

\bibitem[\protect\citeauthoryear{Boley \& Durisen}{Boley \&
  Durisen}{2010}]{boley_2010a}
Boley A.~C.,  Durisen R.~H.,  2010, \mn@doi [ApJ]
  {10.1088/0004-637X/724/1/618}, 724, 618

\bibitem[\protect\citeauthoryear{Boley, Hayfield, Mayer  \& Durisen}{Boley
  et~al.}{2010}]{boley_2010}
Boley A.~C.,  Hayfield T.,  Mayer L.,   Durisen R.~H.,  2010, \mn@doi [Icarus]
  {https://doi.org/10.1016/j.icarus.2010.01.015}, 207, 509

\bibitem[\protect\citeauthoryear{{Boley}, {Helled}  \& {Payne}}{{Boley}
  et~al.}{2011}]{boley_2011}
{Boley} A.~C.,  {Helled} R.,   {Payne} M.~J.,  2011, \mn@doi [\apj]
  {10.1088/0004-637X/735/1/30}, \href
  {http://adsabs.harvard.edu/abs/2011ApJ...735...30B} {735, 30}

\bibitem[\protect\citeauthoryear{{Booth}, {Clarke}, {Madhusudhan}  \&
  {Ilee}}{{Booth} et~al.}{2017}]{booth_2017}
{Booth} R.~A.,  {Clarke} C.~J.,  {Madhusudhan} N.,   {Ilee} J.~D.,  2017,
  preprint, \href {http://adsabs.harvard.edu/abs/2017arXiv170503305B} {}
  (\mn@eprint {arXiv} {1705.03305})

\bibitem[\protect\citeauthoryear{{Boss}}{{Boss}}{1998}]{boss_1998}
{Boss} A.~P.,  1998, \mn@doi [\apj] {10.1086/306036}, \href
  {http://adsabs.harvard.edu/abs/1998ApJ...503..923B} {503, 923}

\bibitem[\protect\citeauthoryear{Brown, Byrne  \& Hindmarsh}{Brown
  et~al.}{1989}]{brown_1989}
Brown P.~N.,  Byrne G.~D.,   Hindmarsh A.~C.,  1989, \mn@doi [SIAM J. Sci.
  Stat. Comput.] {10.1137/0910062}, 10, 1038

\bibitem[\protect\citeauthoryear{{Caselli} \& {Ceccarelli}}{{Caselli} \&
  {Ceccarelli}}{2012}]{caselli_2012}
{Caselli} P.,  {Ceccarelli} C.,  2012, \mn@doi [\aapr]
  {10.1007/s00159-012-0056-x}, \href
  {http://adsabs.harvard.edu/abs/2012A%26ARv..20...56C} {20, 56}

\bibitem[\protect\citeauthoryear{{Cieza} et~al.,}{{Cieza}
  et~al.}{2016}]{cieza_2016}
{Cieza} L.~A.,  et~al., 2016, \mn@doi [\nat] {10.1038/nature18612}, \href
  {http://adsabs.harvard.edu/abs/2016Natur.535..258C} {535, 258}

\bibitem[\protect\citeauthoryear{{Cleeves}, {Bergin}  \& {Harries}}{{Cleeves}
  et~al.}{2015}]{Cleeves_2015}
{Cleeves} L.~I.,  {Bergin} E.~A.,   {Harries} T.~J.,  2015, \mn@doi [\apj]
  {10.1088/0004-637X/807/1/2}, \href
  {http://adsabs.harvard.edu/abs/2015ApJ...807....2C} {807, 2}

\bibitem[\protect\citeauthoryear{{Cossins}, {Lodato}  \& {Clarke}}{{Cossins}
  et~al.}{2010}]{cossins_2010}
{Cossins} P.,  {Lodato} G.,   {Clarke} C.,  2010, \mn@doi [\mnras]
  {10.1111/j.1365-2966.2009.15835.x}, \href
  {http://adsabs.harvard.edu/abs/2010MNRAS.401.2587C} {401, 2587}

\bibitem[\protect\citeauthoryear{{Dipierro}, {Lodato}, {Testi}  \& {de Gregorio
  Monsalvo}}{{Dipierro} et~al.}{2014}]{dipierro_2014}
{Dipierro} G.,  {Lodato} G.,  {Testi} L.,   {de Gregorio Monsalvo} I.,  2014,
  \mn@doi [\mnras] {10.1093/mnras/stu1584}, \href
  {http://adsabs.harvard.edu/abs/2014MNRAS.444.1919D} {444, 1919}

\bibitem[\protect\citeauthoryear{{Dipierro}, {Pinilla}, {Lodato}  \&
  {Testi}}{{Dipierro} et~al.}{2015}]{dipierro_2015}
{Dipierro} G.,  {Pinilla} P.,  {Lodato} G.,   {Testi} L.,  2015, \mn@doi
  [\mnras] {10.1093/mnras/stv970}, \href
  {http://adsabs.harvard.edu/abs/2015MNRAS.451..974D} {451, 974}

\bibitem[\protect\citeauthoryear{{Douglas}, {Caselli}, {Ilee}, {Boley},
  {Hartquist}, {Durisen}  \& {Rawlings}}{{Douglas} et~al.}{2013}]{douglas_2013}
{Douglas} T.~A.,  {Caselli} P.,  {Ilee} J.~D.,  {Boley} A.~C.,  {Hartquist}
  T.~W.,  {Durisen} R.~H.,   {Rawlings} J.~M.~C.,  2013, \mn@doi [\mnras]
  {10.1093/mnras/stt881}, \href
  {http://adsabs.harvard.edu/abs/2013MNRAS.433.2064D} {433, 2064}

\bibitem[\protect\citeauthoryear{{Durisen}, {Boss}, {Mayer}, {Nelson}, {Quinn}
  \& {Rice}}{{Durisen} et~al.}{2007}]{durisen_2007}
{Durisen} R.~H.,  {Boss} A.~P.,  {Mayer} L.,  {Nelson} A.~F.,  {Quinn} T.,
  {Rice} W.~K.~M.,  2007, Protostars and Planets V, \href
  {http://adsabs.harvard.edu/abs/2007prpl.conf..607D} {pp 607--622}

\bibitem[\protect\citeauthoryear{{Dutrey} et~al.,}{{Dutrey}
  et~al.}{2014}]{dutrey_2014}
{Dutrey} A.,  et~al., 2014, \mn@doi [Protostars and Planets VI]
  {10.2458/azu_uapress_9780816531240-ch014}, \href
  {http://adsabs.harvard.edu/abs/2014prpl.conf..317D} {pp 317--338}

\bibitem[\protect\citeauthoryear{{Ehrenfreund} \& {Charnley}}{{Ehrenfreund} \&
  {Charnley}}{2000}]{ehrenfreund_2000}
{Ehrenfreund} P.,  {Charnley} S.~B.,  2000, \mn@doi [\araa]
  {10.1146/annurev.astro.38.1.427}, \href
  {http://adsabs.harvard.edu/abs/2000ARA%26A..38..427E} {38, 427}

\bibitem[\protect\citeauthoryear{{Ehrenfreund} et~al.,}{{Ehrenfreund}
  et~al.}{2003}]{ehrenfreund_2003}
{Ehrenfreund} P.,  et~al., 2003, \mn@doi [\planss]
  {10.1016/S0032-0633(03)00052-7}, \href
  {http://adsabs.harvard.edu/abs/2003P%26SS...51..473E} {51, 473}

\bibitem[\protect\citeauthoryear{{Eisner} \& {Carpenter}}{{Eisner} \&
  {Carpenter}}{2006}]{eisner_2006}
{Eisner} J.~A.,  {Carpenter} J.~M.,  2006, \mn@doi [\apj] {10.1086/500637},
  \href {http://adsabs.harvard.edu/abs/2006ApJ...641.1162E} {641, 1162}

\bibitem[\protect\citeauthoryear{{Eistrup}, {Walsh}  \& {van
  Dishoeck}}{{Eistrup} et~al.}{2016}]{eistrup_2016}
{Eistrup} C.,  {Walsh} C.,   {van Dishoeck} E.~F.,  2016, \mn@doi [\aap]
  {10.1051/0004-6361/201628509}, \href
  {http://ukads.nottingham.ac.uk/abs/2016A%26A...595A..83E} {595, A83}

\bibitem[\protect\citeauthoryear{{Evans}, {Ilee}, {Boley}, {Caselli},
  {Durisen}, {Hartquist}  \& {Rawlings}}{{Evans} et~al.}{2015}]{evans_2015}
{Evans} M.~G.,  {Ilee} J.~D.,  {Boley} A.~C.,  {Caselli} P.,  {Durisen} R.~H.,
  {Hartquist} T.~W.,   {Rawlings} J.~M.~C.,  2015, \mn@doi [\mnras]
  {10.1093/mnras/stv1698}, \href
  {http://adsabs.harvard.edu/abs/2015MNRAS.453.1147E} {453, 1147}

\bibitem[\protect\citeauthoryear{{Evans} et~al.,}{{Evans}
  et~al.}{2017}]{evans_2017}
{Evans} M.~G.,  et~al., 2017, preprint, \href
  {http://adsabs.harvard.edu/abs/2017arXiv170600254E} {} (\mn@eprint {arXiv}
  {1706.00254})

\bibitem[\protect\citeauthoryear{{Factor} et~al.,}{{Factor}
  et~al.}{2017}]{factor_2017}
{Factor} S.~M.,  et~al., 2017, \mn@doi [\aj] {10.3847/1538-3881/aa6c2c}, \href
  {http://adsabs.harvard.edu/abs/2017AJ....153..233F} {153, 233}

\bibitem[\protect\citeauthoryear{{Forgan} \& {Rice}}{{Forgan} \&
  {Rice}}{2013}]{forgan_2013}
{Forgan} D.,  {Rice} K.,  2013, \mn@doi [\mnras] {10.1093/mnras/stt672}, \href
  {http://adsabs.harvard.edu/abs/2013MNRAS.432.3168F} {432, 3168}

\bibitem[\protect\citeauthoryear{Forgan, Rice, Stamatellos  \&
  Whitworth}{Forgan et~al.}{2009}]{forgan_2009}
Forgan D.~H.,  Rice K.,  Stamatellos D.,   Whitworth A.~P.,  2009, \mn@doi
  [MNRAS] {10.1111/j.1365-2966.2008.14373.x}, 394, 882

\bibitem[\protect\citeauthoryear{{Forgan}, {Rice}, {Cossins}  \&
  {Lodato}}{{Forgan} et~al.}{2011}]{forgan_2011}
{Forgan} D.,  {Rice} K.,  {Cossins} P.,   {Lodato} G.,  2011, \mn@doi [\mnras]
  {10.1111/j.1365-2966.2010.17500.x}, \href
  {http://adsabs.harvard.edu/abs/2011MNRAS.410..994F} {410, 994}

\bibitem[\protect\citeauthoryear{{Forgan}, {Parker}  \& {Rice}}{{Forgan}
  et~al.}{2015}]{forgan_2015}
{Forgan} D.,  {Parker} R.~J.,   {Rice} K.,  2015, \mn@doi [\mnras]
  {10.1093/mnras/stu2504}, \href
  {http://adsabs.harvard.edu/abs/2015MNRAS.447..836F} {447, 836}

\bibitem[\protect\citeauthoryear{{Frimann} et~al.,}{{Frimann}
  et~al.}{2017}]{frimann_2017}
{Frimann} S.,  et~al., 2017, preprint, \href
  {http://adsabs.harvard.edu/abs/2017arXiv170310225F} {} (\mn@eprint {arXiv}
  {1703.10225})

\bibitem[\protect\citeauthoryear{{Gammie}}{{Gammie}}{2001}]{gammie_2001}
{Gammie} C.~F.,  2001, \mn@doi [\apj] {10.1086/320631}, \href
  {http://adsabs.harvard.edu/abs/2001ApJ...553..174G} {553, 174}

\bibitem[\protect\citeauthoryear{{Hall}, {Forgan}, {Rice}, {Harries},
  {Klaassen}  \& {Biller}}{{Hall} et~al.}{2016}]{hall_2016}
{Hall} C.,  {Forgan} D.,  {Rice} K.,  {Harries} T.~J.,  {Klaassen} P.~D.,
  {Biller} B.,  2016, \mn@doi [\mnras] {10.1093/mnras/stw296}, \href
  {http://adsabs.harvard.edu/abs/2016MNRAS.458..306H} {458, 306}

\bibitem[\protect\citeauthoryear{{Hall}, {Forgan}  \& {Rice}}{{Hall}
  et~al.}{2017}]{hall_2017}
{Hall} C.,  {Forgan} D.,   {Rice} K.,  2017, preprint, \href
  {http://adsabs.harvard.edu/abs/2017arXiv170506690H} {} (\mn@eprint {arXiv}
  {1705.06690})

\bibitem[\protect\citeauthoryear{{Haworth} et~al.,}{{Haworth}
  et~al.}{2016}]{haworth_2016}
{Haworth} T.~J.,  et~al., 2016, \mn@doi [\pasa] {10.1017/pasa.2016.45}, \href
  {http://adsabs.harvard.edu/abs/2016PASA...33...53H} {33, e053}

\bibitem[\protect\citeauthoryear{{Heiter}}{{Heiter}}{2002}]{heiter_2002}
{Heiter} U.,  2002, \mn@doi [\aap] {10.1051/0004-6361:20011593}, \href
  {http://adsabs.harvard.edu/abs/2002A%26A...381..959H} {381, 959}

\bibitem[\protect\citeauthoryear{{Helled}, {Podolak}  \& {Kovetz}}{{Helled}
  et~al.}{2006}]{helled_2006}
{Helled} R.,  {Podolak} M.,   {Kovetz} A.,  2006, \mn@doi [\icarus]
  {10.1016/j.icarus.2006.06.011}, \href
  {http://adsabs.harvard.edu/abs/2006Icar..185...64H} {185, 64}

\bibitem[\protect\citeauthoryear{{Helled} et~al.,}{{Helled}
  et~al.}{2014}]{helled_2014}
{Helled} R.,  et~al., 2014, \mn@doi [Protostars and Planets VI]
  {10.2458/azu_uapress_9780816531240-ch028}, \href
  {http://adsabs.harvard.edu/abs/2014prpl.conf..643H} {pp 643--665}

\bibitem[\protect\citeauthoryear{{Henning} \& {Semenov}}{{Henning} \&
  {Semenov}}{2013}]{henning_2013}
{Henning} T.,  {Semenov} D.,  2013, \mn@doi [Chemical Reviews]
  {10.1021/cr400128p}, \href
  {http://adsabs.harvard.edu/abs/2013ChRv..113.9016H} {113, 9016}

\bibitem[\protect\citeauthoryear{{Hincelin}, {Wakelam}, {Commer{\c c}on},
  {Hersant}  \& {Guilloteau}}{{Hincelin} et~al.}{2013}]{hincelin_2013}
{Hincelin} U.,  {Wakelam} V.,  {Commer{\c c}on} B.,  {Hersant} F.,
  {Guilloteau} S.,  2013, \mn@doi [\apj] {10.1088/0004-637X/775/1/44}, \href
  {http://adsabs.harvard.edu/abs/2013ApJ...775...44H} {775, 44}

\bibitem[\protect\citeauthoryear{{Ilee}, {Boley}, {Caselli}, {Durisen},
  {Hartquist}  \& {Rawlings}}{{Ilee} et~al.}{2011}]{ilee_2011}
{Ilee} J.~D.,  {Boley} A.~C.,  {Caselli} P.,  {Durisen} R.~H.,  {Hartquist}
  T.~W.,   {Rawlings} J.~M.~C.,  2011, \mn@doi [\mnras]
  {10.1111/j.1365-2966.2011.19455.x}, \href
  {http://adsabs.harvard.edu/abs/2011MNRAS.417.2950I} {417, 2950}

\bibitem[\protect\citeauthoryear{{Joos}, {Hennebelle}  \& {Ciardi}}{{Joos}
  et~al.}{2012}]{joos_2012}
{Joos} M.,  {Hennebelle} P.,   {Ciardi} A.,  2012, \mn@doi [\aap]
  {10.1051/0004-6361/201118730}, \href
  {http://adsabs.harvard.edu/abs/2012A%26A...543A.128J} {543, A128}

\bibitem[\protect\citeauthoryear{{Khajenabi}, {Kazrani}  \&
  {Shadmehri}}{{Khajenabi} et~al.}{2017}]{khajenabi_2017}
{Khajenabi} F.,  {Kazrani} K.,   {Shadmehri} M.,  2017, preprint, \href
  {http://adsabs.harvard.edu/abs/2017arXiv170500288K} {} (\mn@eprint {arXiv}
  {1705.00288})

\bibitem[\protect\citeauthoryear{{Lodato} \& {Rice}}{{Lodato} \&
  {Rice}}{2004}]{lodato_2004}
{Lodato} G.,  {Rice} W.~K.~M.,  2004, \mn@doi [\mnras]
  {10.1111/j.1365-2966.2004.07811.x}, \href
  {http://adsabs.harvard.edu/abs/2004MNRAS.351..630L} {351, 630}

\bibitem[\protect\citeauthoryear{{Madhusudhan}}{{Madhusudhan}}{2012}]{madhusudhan_2012}
{Madhusudhan} N.,  2012, \mn@doi [\apj] {10.1088/0004-637X/758/1/36}, \href
  {http://adsabs.harvard.edu/abs/2012ApJ...758...36M} {758, 36}

\bibitem[\protect\citeauthoryear{{Madhusudhan}, {Amin}  \&
  {Kennedy}}{{Madhusudhan} et~al.}{2014}]{madhusudhan_2014}
{Madhusudhan} N.,  {Amin} M.~A.,   {Kennedy} G.~M.,  2014, \mn@doi [\apjl]
  {10.1088/2041-8205/794/1/L12}, \href
  {http://adsabs.harvard.edu/abs/2014ApJ...794L..12M} {794, L12}

\bibitem[\protect\citeauthoryear{{Marois}, {Macintosh}, {Barman}, {Zuckerman},
  {Song}, {Patience}, {Lafreni{\`e}re}  \& {Doyon}}{{Marois}
  et~al.}{2008}]{marois_2008}
{Marois} C.,  {Macintosh} B.,  {Barman} T.,  {Zuckerman} B.,  {Song} I.,
  {Patience} J.,  {Lafreni{\`e}re} D.,   {Doyon} R.,  2008, \mn@doi [Science]
  {10.1126/science.1166585}, \href
  {http://adsabs.harvard.edu/abs/2008Sci...322.1348M} {322, 1348}

\bibitem[\protect\citeauthoryear{Mayer, Lufkin, Quinn  \& Wadsley}{Mayer
  et~al.}{2007}]{mayer_2007}
Mayer L.,  Lufkin G.,  Quinn T.,   Wadsley J.,  2007, \mn@doi [ApJ]
  {10.1086/518433}, 661, L77

\bibitem[\protect\citeauthoryear{Meru \& Bate}{Meru \& Bate}{2011}]{meru_2011}
Meru F.,  Bate M.~R.,  2011, \mn@doi [MNRAS]
  {10.1111/j.1745-3933.2010.00978.x}, 411, L1

\bibitem[\protect\citeauthoryear{{Meru}, {Juh{\'a}sz}, {Ilee}, {Clarke},
  {Rosotti}  \& {Booth}}{{Meru} et~al.}{2017}]{Meru_2017}
{Meru} F.,  {Juh{\'a}sz} A.,  {Ilee} J.~D.,  {Clarke} C.~J.,  {Rosotti} G.~P.,
   {Booth} R.~A.,  2017, \mn@doi [\apjl] {10.3847/2041-8213/aa6837}, \href
  {http://adsabs.harvard.edu/abs/2017ApJ...839L..24M} {839, L24}

\bibitem[\protect\citeauthoryear{{Millar}, {Farquhar}  \& {Willacy}}{{Millar}
  et~al.}{1997}]{millar_1997}
{Millar} T.~J.,  {Farquhar} P.~R.~A.,   {Willacy} K.,  1997, \mn@doi [\aaps]
  {10.1051/aas:1997118}, \href
  {http://adsabs.harvard.edu/abs/1997A%26AS..121..139M} {121}

\bibitem[\protect\citeauthoryear{{Narayanan}, {Kulesa}, {Boss}  \&
  {Walker}}{{Narayanan} et~al.}{2006}]{narayanan_2006}
{Narayanan} D.,  {Kulesa} C.~A.,  {Boss} A.,   {Walker} C.~K.,  2006, \mn@doi
  [\apj] {10.1086/505619}, \href
  {http://adsabs.harvard.edu/abs/2006ApJ...647.1426N} {647, 1426}

\bibitem[\protect\citeauthoryear{{Nayakshin}}{{Nayakshin}}{2010a}]{nayakshin_2010a}
{Nayakshin} S.,  2010a, \mn@doi [\mnras] {10.1111/j.1745-3933.2010.00923.x},
  \href {http://adsabs.harvard.edu/abs/2010MNRAS.408L..36N} {408, L36}

\bibitem[\protect\citeauthoryear{{Nayakshin}}{{Nayakshin}}{2010b}]{nayakshin_2010b}
{Nayakshin} S.,  2010b, \mn@doi [\mnras] {10.1111/j.1365-2966.2010.17289.x},
  \href {http://adsabs.harvard.edu/abs/2010MNRAS.408.2381N} {408, 2381}

\bibitem[\protect\citeauthoryear{{Nayakshin} \& {Fletcher}}{{Nayakshin} \&
  {Fletcher}}{2015}]{nayakshin_2015}
{Nayakshin} S.,  {Fletcher} M.,  2015, \mn@doi [\mnras]
  {10.1093/mnras/stv1354}, \href
  {http://adsabs.harvard.edu/abs/2015MNRAS.452.1654N} {452, 1654}

\bibitem[\protect\citeauthoryear{{{\"O}berg}, {Murray-Clay}  \&
  {Bergin}}{{{\"O}berg} et~al.}{2011}]{oberg_2011}
{{\"O}berg} K.~I.,  {Murray-Clay} R.,   {Bergin} E.~A.,  2011, \mn@doi [\apjl]
  {10.1088/2041-8205/743/1/L16}, \href
  {http://adsabs.harvard.edu/abs/2011ApJ...743L..16O} {743, L16}

\bibitem[\protect\citeauthoryear{{Paardekooper}}{{Paardekooper}}{2012}]{paardekooper_2012}
{Paardekooper} S.-J.,  2012, \mn@doi [\mnras]
  {10.1111/j.1365-2966.2012.20553.x}, \href
  {http://adsabs.harvard.edu/abs/2012MNRAS.421.3286P} {421, 3286}

\bibitem[\protect\citeauthoryear{{Pani{\'c}} \& {Min}}{{Pani{\'c}} \&
  {Min}}{2017}]{panic_2017}
{Pani{\'c}} O.,  {Min} M.,  2017, \mn@doi [\mnras] {10.1093/mnras/stx114},
  \href {http://adsabs.harvard.edu/abs/2017MNRAS.tmp..115P} {}

\bibitem[\protect\citeauthoryear{P{\'e}rez et~al.,}{P{\'e}rez
  et~al.}{2016}]{Perez_2016}
P{\'e}rez L.~M.,  et~al., 2016, \mn@doi [Science] {10.1126/science.aaf8296},
  353, 1519

\bibitem[\protect\citeauthoryear{{Pollack}, {Hubickyj}, {Bodenheimer},
  {Lissauer}, {Podolak}  \& {Greenzweig}}{{Pollack}
  et~al.}{1996}]{pollack_1996}
{Pollack} J.~B.,  {Hubickyj} O.,  {Bodenheimer} P.,  {Lissauer} J.~J.,
  {Podolak} M.,   {Greenzweig} Y.,  1996, \mn@doi [\icarus]
  {10.1006/icar.1996.0190}, \href
  {http://adsabs.harvard.edu/abs/1996Icar..124...62P} {124, 62}

\bibitem[\protect\citeauthoryear{{Price}}{{Price}}{2007}]{price_2007}
{Price} D.~J.,  2007, \mn@doi [\pasa] {10.1071/AS07022}, \href
  {http://adsabs.harvard.edu/abs/2007PASA...24..159P} {24, 159}

\bibitem[\protect\citeauthoryear{{Qi} et~al.,}{{Qi} et~al.}{2013}]{qi_2013}
{Qi} C.,  et~al., 2013, \mn@doi [Science] {10.1126/science.1239560}, \href
  {http://adsabs.harvard.edu/abs/2013Sci...341..630Q} {341, 630}

\bibitem[\protect\citeauthoryear{{Rice}, {Lodato}  \& {Armitage}}{{Rice}
  et~al.}{2005}]{rice_2005}
{Rice} W.~K.~M.,  {Lodato} G.,   {Armitage} P.~J.,  2005, \mn@doi [\mnras]
  {10.1111/j.1745-3933.2005.00105.x}, \href
  {http://adsabs.harvard.edu/abs/2005MNRAS.364L..56R} {364, L56}

\bibitem[\protect\citeauthoryear{{Rice}, {Mayo}  \& {Armitage}}{{Rice}
  et~al.}{2010}]{rice_2010}
{Rice} W.~K.~M.,  {Mayo} J.~H.,   {Armitage} P.~J.,  2010, \mn@doi [\mnras]
  {10.1111/j.1365-2966.2009.15992.x}, \href
  {http://adsabs.harvard.edu/abs/2010MNRAS.402.1740R} {402, 1740}

\bibitem[\protect\citeauthoryear{{Saumon} \& {Guillot}}{{Saumon} \&
  {Guillot}}{2004}]{saumon_2004}
{Saumon} D.,  {Guillot} T.,  2004, \mn@doi [\apj] {10.1086/421257}, \href
  {http://adsabs.harvard.edu/abs/2004ApJ...609.1170S} {609, 1170}

\bibitem[\protect\citeauthoryear{Stamatellos, Whitworth, Bisbas  \&
  Goodwin}{Stamatellos et~al.}{2007}]{stamatellos_2007a}
Stamatellos D.,  Whitworth A.~P.,  Bisbas T.,   Goodwin S.,  2007, \mn@doi
  [A{\&}A] {10.1051/0004-6361:20077373}, 475, 37

\bibitem[\protect\citeauthoryear{{Tobin} et~al.,}{{Tobin}
  et~al.}{2015}]{tobin_2015}
{Tobin} J.~J.,  et~al., 2015, \mn@doi [\apj] {10.1088/0004-637X/805/2/125},
  \href {http://adsabs.harvard.edu/abs/2015ApJ...805..125T} {805, 125}

\bibitem[\protect\citeauthoryear{{Tobin} et~al.,}{{Tobin}
  et~al.}{2016}]{Tobin_2016}
{Tobin} J.~J.,  et~al., 2016, \mn@doi [\nat] {10.1038/nature20094}, \href
  {http://adsabs.harvard.edu/abs/2016Natur.538..483T} {538, 483}

\bibitem[\protect\citeauthoryear{{Toomre}}{{Toomre}}{1964}]{toomre_1964}
{Toomre} A.,  1964, \mn@doi [\apj] {10.1086/147861}, \href
  {http://adsabs.harvard.edu/abs/1964ApJ...139.1217T} {139, 1217}

\bibitem[\protect\citeauthoryear{{Vigan} et~al.,}{{Vigan}
  et~al.}{2017}]{Vigan2017}
{Vigan} A.,  et~al., 2017, preprint, \href
  {http://adsabs.harvard.edu/abs/2017arXiv170305322V} {} (\mn@eprint {arXiv}
  {1703.05322})

\bibitem[\protect\citeauthoryear{{Wakelam} et~al.,}{{Wakelam}
  et~al.}{2012}]{wakelam_2012}
{Wakelam} V.,  et~al., 2012, \mn@doi [\apjs] {10.1088/0067-0049/199/1/21},
  \href {http://adsabs.harvard.edu/abs/2012ApJS..199...21W} {199, 21}

\bibitem[\protect\citeauthoryear{{Wang}, {Bell}, {Iedema}, {Tsekouras}  \&
  {Cowin}}{{Wang} et~al.}{2005}]{wang_2005}
{Wang} H.,  {Bell} R.~C.,  {Iedema} M.~J.,  {Tsekouras} A.~A.,   {Cowin} J.~P.,
   2005, \mn@doi [\apj] {10.1086/427072}, \href
  {http://adsabs.harvard.edu/abs/2005ApJ...620.1027W} {620, 1027}

\bibitem[\protect\citeauthoryear{Whitehouse \& Bate}{Whitehouse \&
  Bate}{2004}]{whitehouse_2004}
Whitehouse S.~C.,  Bate M.~R.,  2004, \mn@doi [MNRAS]
  {10.1111/j.1365-2966.2004.08131.x}, 353, 1078

\bibitem[\protect\citeauthoryear{{Young} \& {Clarke}}{{Young} \&
  {Clarke}}{2016}]{young_2016}
{Young} M.~D.,  {Clarke} C.~J.,  2016, \mn@doi [\mnras]
  {10.1093/mnras/stv2378}, \href
  {http://adsabs.harvard.edu/abs/2016MNRAS.455.1438Y} {455, 1438}

\makeatother
\end{thebibliography}

\appendix

\section{Physical \& chemical structure of the fragments}
\label{app:frag}

\begin{figure*}
\includegraphics[width=\textwidth]{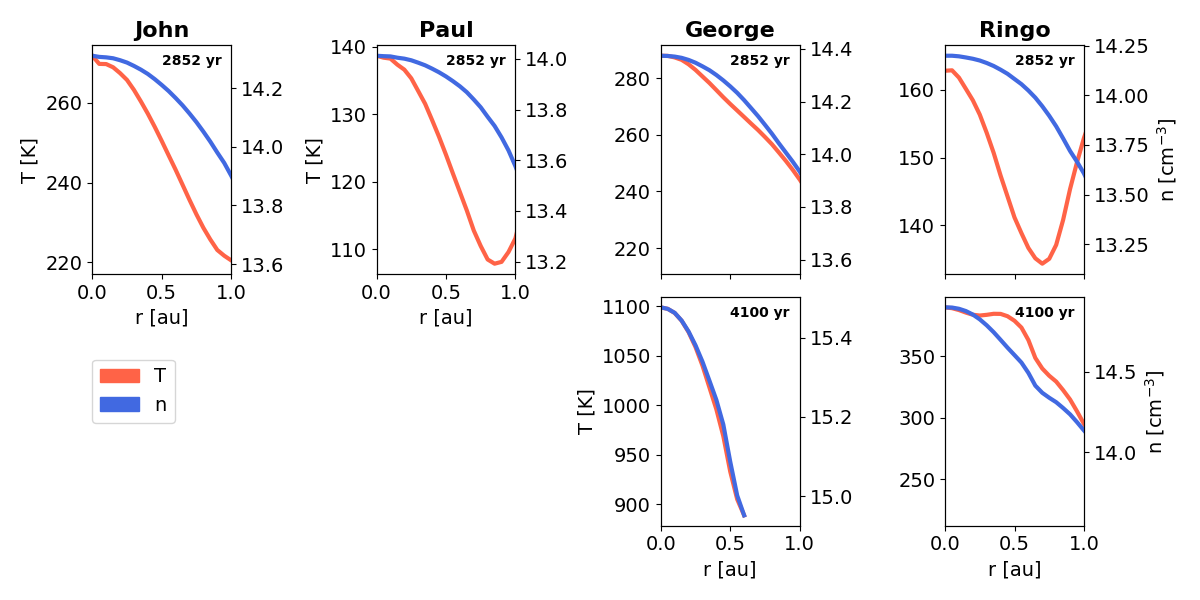} 
\caption{Spherically averaged physical conditions towards the fragments as a function of time.  The upper row shows conditions at 2852\,yr and the lower row at 4100\,yr.  Upward turns exhibited by some of the temperature profiles are due to the inclusion of particles at higher $z$ in the disc, which experience somewhat higher temperatures.}
\label{fig:fragphys}
\end{figure*}

Figure \ref{fig:fragphys} shows the averaged physical and chemical structure of the fragments are various times throughout the simulation.

\begin{figure*}
\includegraphics[width=\textwidth]{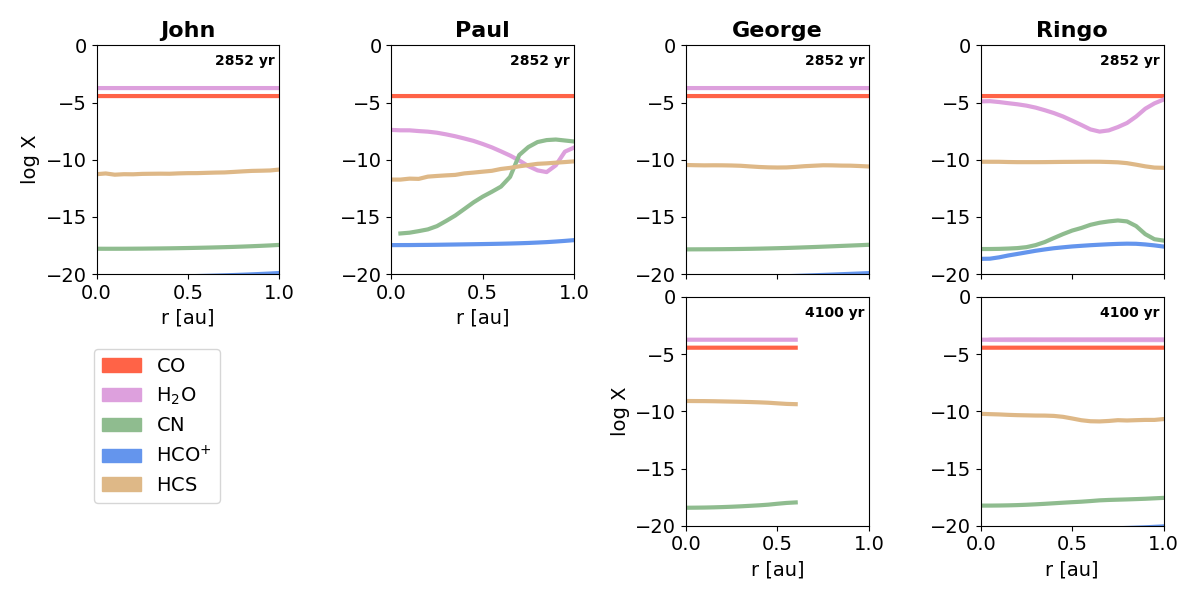} 
\caption{Spherically averaged fractional aboundances for selected species towards the fragments as a function of time.  The upper row shows abundances at 2852\,yr and the lower row at 4100\,yr.}
\label{fig:fragchem}
\end{figure*}

\section{Column density images for all species}
\label{app:chem}

\begin{figure*}
\includegraphics[width=\textwidth]{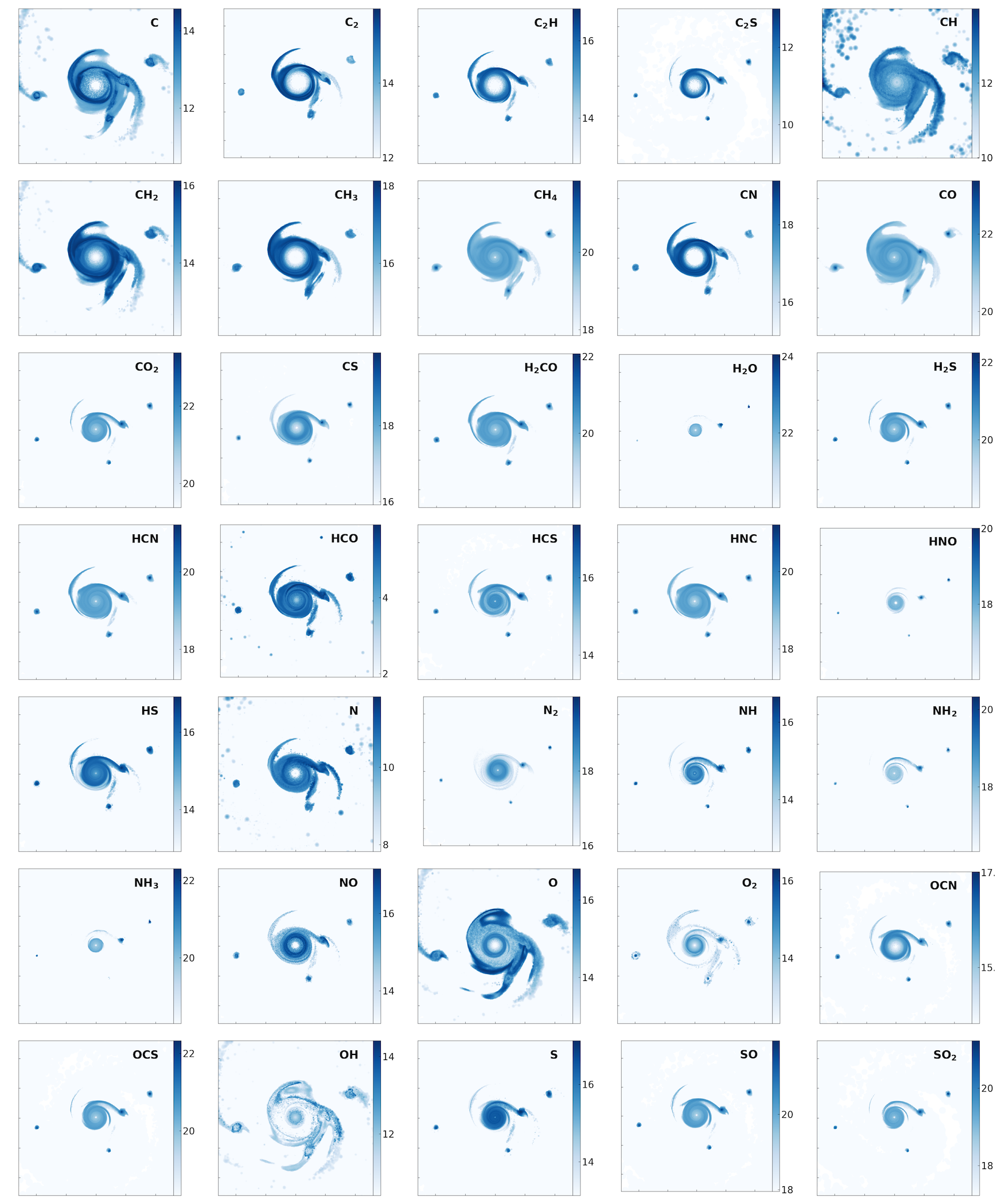} 
\caption{The dominant neutral gas phase species at $t=2852$\,yr.  Panels span 130\,au and the colour bar gives the logarithm of the column density in cm$^{-2}$.}
\label{fig:all_gas}
\end{figure*}

\begin{figure*}
\includegraphics[width=\textwidth]{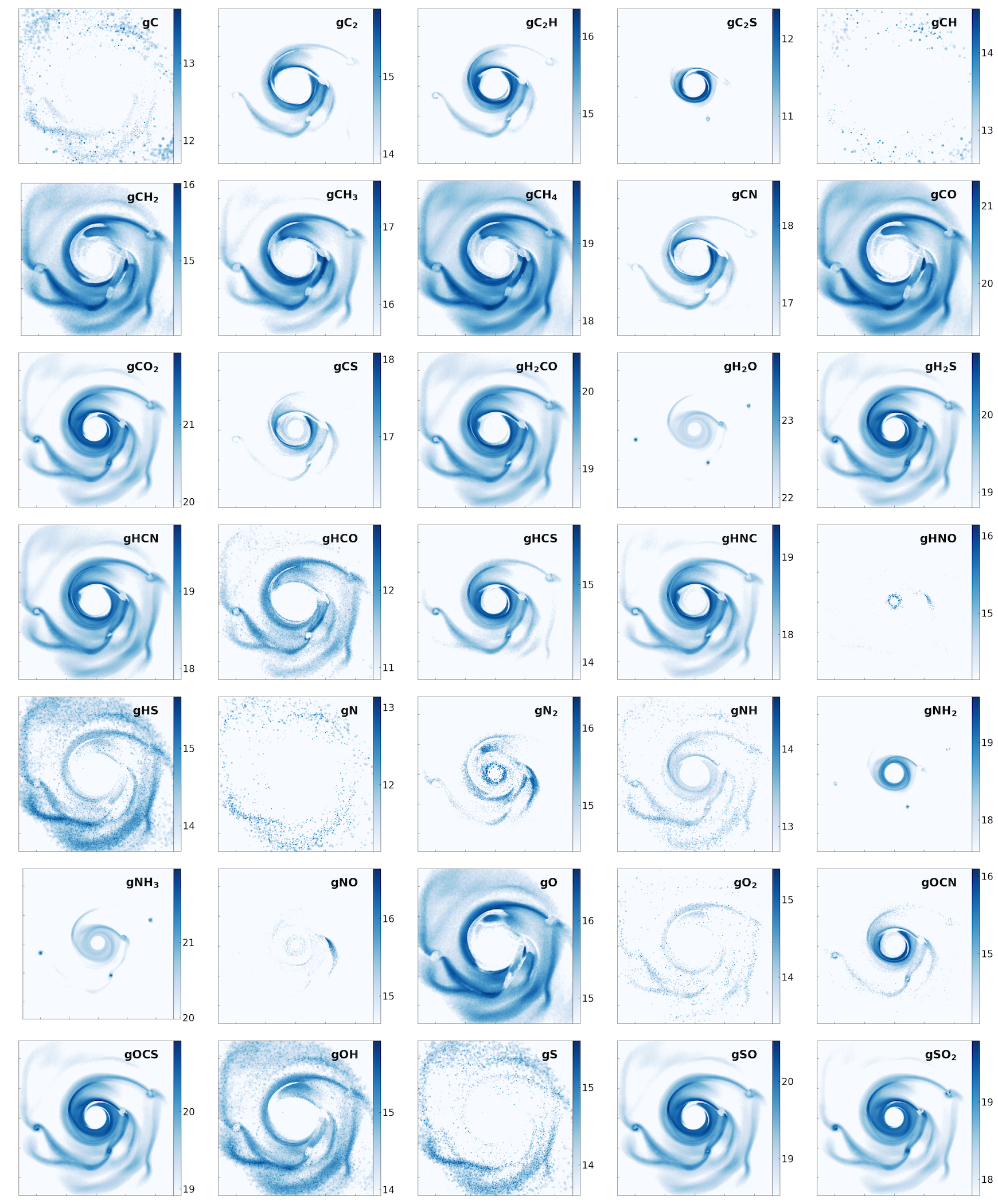} 
\caption{The dominant grain surface species at $t=2852$\,yr.  Panels span 130\,au and the colour bar gives the logarithm of the column density in cm$^{-2}$.}
\label{fig:all_grain}
\end{figure*}

\begin{figure*}
\includegraphics[width=\textwidth]{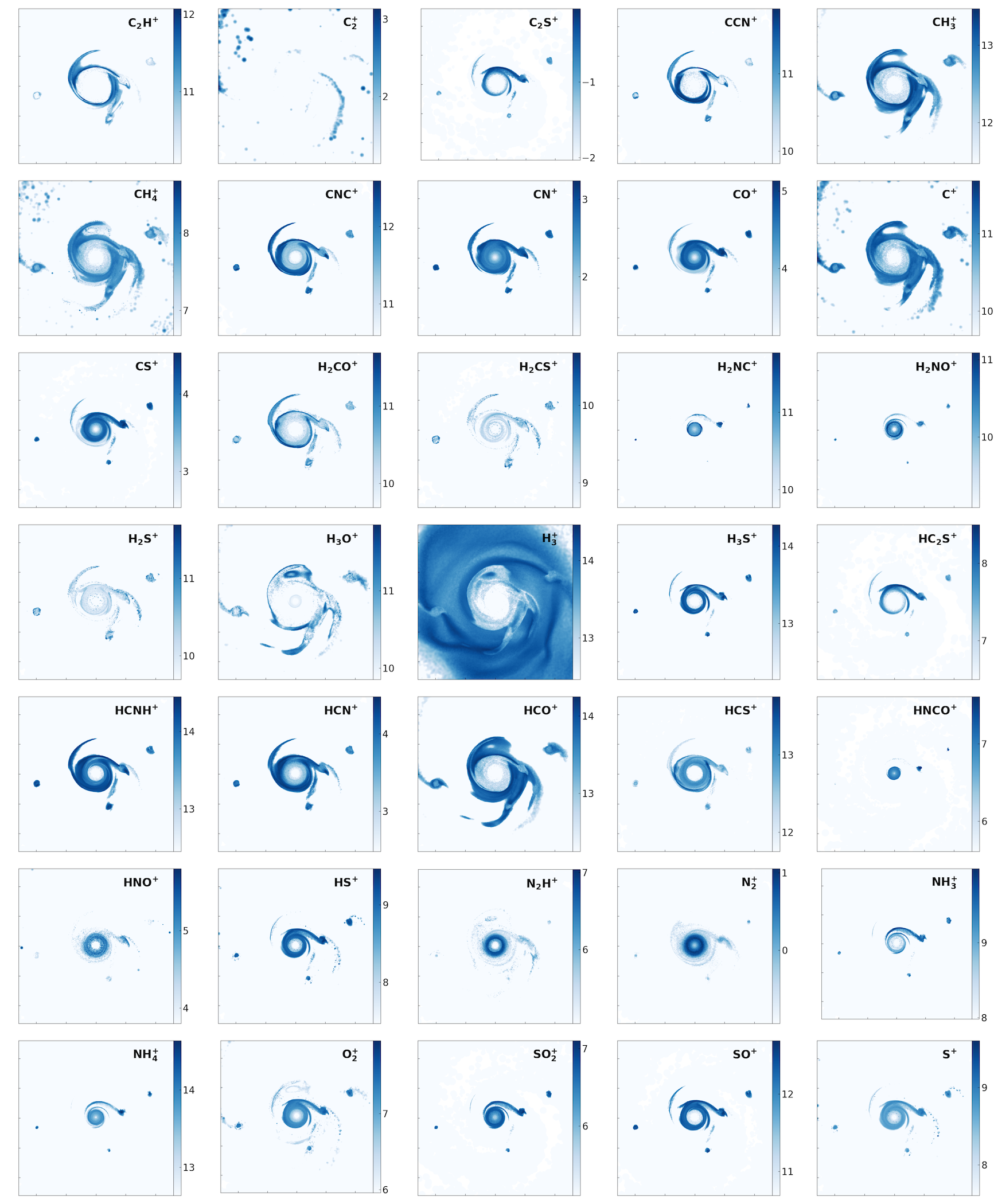} 
\caption{The dominant cations in the gas phase at $t=2852$\,yr.  Panels span 130\,au and the colour bar gives the logarithm of the column density in cm$^{-2}$.}
\label{fig:all_ions}
\end{figure*}

Figures \ref{fig:all_gas}, \ref{fig:all_grain} and \ref{fig:all_ions} show the column density of the dominant neutral gas phase species, grain surface species, and cations, respectively.     

\bsp	
\label{lastpage}
\end{document}